\documentclass{tran-l}

\usepackage{xcolor}
\definecolor{myblue}{RGB}{0, 63, 0}
\usepackage{mathrsfs}
\usepackage{soul}
\theoremstyle{definition}

\theoremstyle{remark}

\numberwithin{equation}{section}
\newcommand{\dd}{\mathrm{d}}

\usepackage{graphicx}
\begin{document}

\title{Chaos, Ito-Stratonovich dilemma, and topological supersymmetry}

\author{Igor V. Ovchinnikov}
\address{R\&D, CSD, ThermoFisher Scientific Inc, 200 Oyster Point, South San Francisco, 94080, CA, USA}
\email{igor.vlad.ovchinnikov@gmail.com}
\thanks{This work was partly conducted at the Device Research Laboratory, Department of Electrical Engineering, University of California at Los Angeles, Los Angeles, CA 90095, USA}



\begin{abstract} 
It was recently established that the formalism of the generalized transfer operator (GTO) of dynamical systems (DS) theory, applied to stochastic differential equations (SDEs) of arbitrary form, belongs to the family of cohomological topological field theories (TFT) -- a class of models at the intersection of algebraic topology and high-energy physics. This interdisciplinary approach, which can be called the supersymmetric theory of stochastic dynamics (STS), can be seen as an algebraic dual to the traditional set-theoretic framework of the DS theory, with its algebraic structure enabling the extension of some DS theory concepts to stochastic dynamics. Moreover, it reveals the presence of a topological supersymmetry (TS) in the GTOs of all SDEs. It also shows that among the various definitions of chaos, positive "pressure", defined as the logarithm of the GTO spectral radius, stands out as particularly meaningful from a physical perspective, as it corresponds to the spontaneous breakdown of TS on the TFT side. Via the Goldstone theorem, this definition has a potential to provide the long-sought explanation for the experimental signature of chaotic dynamics known as 1/f noise. Additionally, STS clarifies that among the various existing interpretations of SDEs, only the Stratonovich interpretation yields evolution operators that match the corresponding GTOs and, consequently, have a clear-cut mathematical meaning. Here, we discuss these and other aspects of STS from both the DS theory and TFT perspectives, focusing on links between these two fields and providing mathematical concepts with physical interpretations that may be useful in some contexts.
\end{abstract}

\maketitle

\section{Introduction}
Originally introduced as an extension of the theory of elementary particles, \cite{KaneShifman2000} supersymmetry has since evolved into a mathematical concept \cite{Wit82} that underlies cohomological topological field theories (TFTs), \cite{Baulieu_1988,Witten88,Witten881,Baulieu_1989,TFT_BOOK,labastida1989} a family of models that bridge algebraic topology and high-energy physics. One interesting member of this family \cite{Baulieu_1988,Gozzi3} is the Parisi-Sourlas approach to Langevin stochastic differential equations (SDEs) \cite{ParSour1,ParSour,Lyapunov_SUSY} and its extensions to other classes of SDEs, \cite{Gozzi3,Niemi2,Kurchan,KS} including general-form SDEs \cite{OVCHINNIKOV2024114611} capable of exhibiting chaos -- a ubiquitous phenomenon with a long history in science \cite{Shep14,Rue14,Mot14,Chaos_orig} and a central topic in dynamical systems (DS) theory. \cite{Handbook_of_DS,Review_Top_Entropy} In this way, this framework, that can be called supersymmetric theory of stochastic dynamics (STS), connects TFTs with DS theory, offering interdisciplinary insights that may enrich both fields.
    
From a DS theory perspective, an interesting insight from STS is that one of the definitions of chaos -- the emergence of positive "pressure" \cite{Rue02,Rue1990} -- is equivalent to the spontaneous breakdown of topological supersymmetry (TS), an inherent property of all stochastic DSs. This presents two reasons why this definition stands out among other possible ways to define chaos. First, it makes very good physical sense -- Richard Feynman might not have called turbulence "the most important unsolved problem of classical physics" had he been aware that (hydrodynamic) chaos {belongs to the most numerous family of qualitative physical phenomena that arise from spontaneous breakdown of various symmetries of nature.} Second, through the Goldstone theorem, spontaneous TS breaking picture of chaos may provide a long-sought explanation for the experimental signature of chaos, known as 1/f noise. \cite{Keshner_1_f_noise_1982,RevModPhys.53.497,Asc11,PhysRevLett.97.118102}

From the perspective of the TFT approach to SDEs, STS sheds new light on the operator ordering conventions in stochastic evolution operators which, in traditional theory of SDEs, \cite{Kunita2019,Stochastic_differential_geometry_at} correspond to different coexisting interpretations of stochastic dynamics. It shows that only the Stratonovich interpretation yields stochastic evolution operators that match the generalized transfer operators of the DS theory, which are unique and have a very natural mathematical meaning.

Over time, DS theory and TFTs have developed distinct perspectives on concepts that overlap within STS. Relating these perspectives may help facilitate interdisciplinary exchange. To this end, we present two complementary discussions of STS, each drawing connections to the other, and provide physical interpretations of mathematical concepts that may help strengthening the links between the two fields. In Sec.\ref{Sec:Dynamical_Systems_Theory}, we discuss continuous-time stochastic DSs. In Sec.\ref{Sec:Pathintegral_Representation_of_Stochastic_Dynamics}, we examine the conventional approach to SDEs and their TFT representation. In Sec.\ref{Sec:STS_TFT}, we focus on the topological aspects of STS such as a close relation between instantons and Morse-Smale DSs. Sec.\ref{Sec:Phase_Diagram} offers a qualitative discussion of the STS perspective on 1/f noise and the "edge of chaos." We conclude in Sec. \ref{Sec:Conclusion}.

\section{Continuous-time stochastic dynamical systems}
\label{Sec:Dynamical_Systems_Theory}
One of the primary objects of interest in DS theory is a continuous-time deterministic DS, \emph{i.e.}, an ordinary differential equation (ODE), $\dot x(t) = F(x(t))$, where $x\in X$ is a point in the phase (or state) space, $X$, which, for concreteness, can be assumed to be a closed smooth manifold, and the law of evolution is represented by a sufficiently smooth flow vector field, $F\in TX$, from the tangent space of $X$.

Deterministic dynamics is a mathematical idealization, as real-world DSs are inevitably subject to unpredictable environmental influence called noise. A more general formulation that incorporates the influence from the noise is given by the following non-autonomous extension of the continuous-time dynamics,\footnote{Here and in the following, the summation over repeated indices is assumed. Moreover, to prevent excessive notation, the vector indices are suppressed so that formulas appear as if the phase space was 1D. To avoid confusion, the vector indices are given explicitly in some formulas.}
\begin{eqnarray}
\dot x(t) = F(x(t))+(2\Theta)^{1/2}G_a(x(t))\xi^a(t)\equiv{\mathscr F}(x(t),\xi(t)), \label{SDE}   
\end{eqnarray}
where $G_a \in TX, a=1, \ldots, D, D=\dim X$ is a set of sufficiently smooth vector fields that specify how the DS is coupled to the time-dependent noise, $\xi(t)\in\mathbb{R}^D$.\footnote{In the literature, the noise is called additive/multiplicative depending on whether $G_a$'s are independent/dependent on $x$} 

\begin{figure}[t]
    \centering
    \includegraphics[height=3.9cm,width=3.6cm]{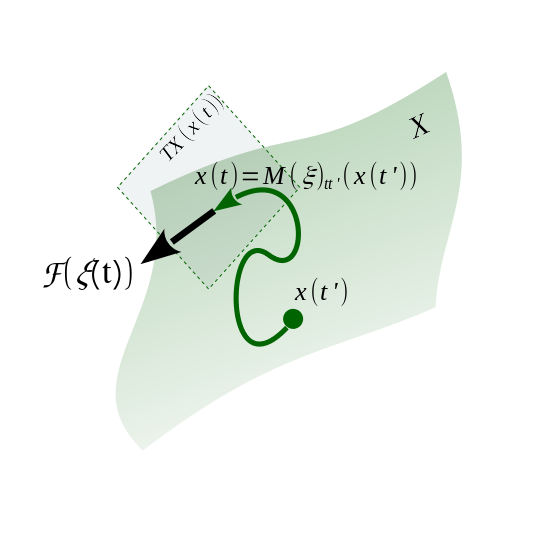}\includegraphics[width=0.55\linewidth,height=3.5cm]{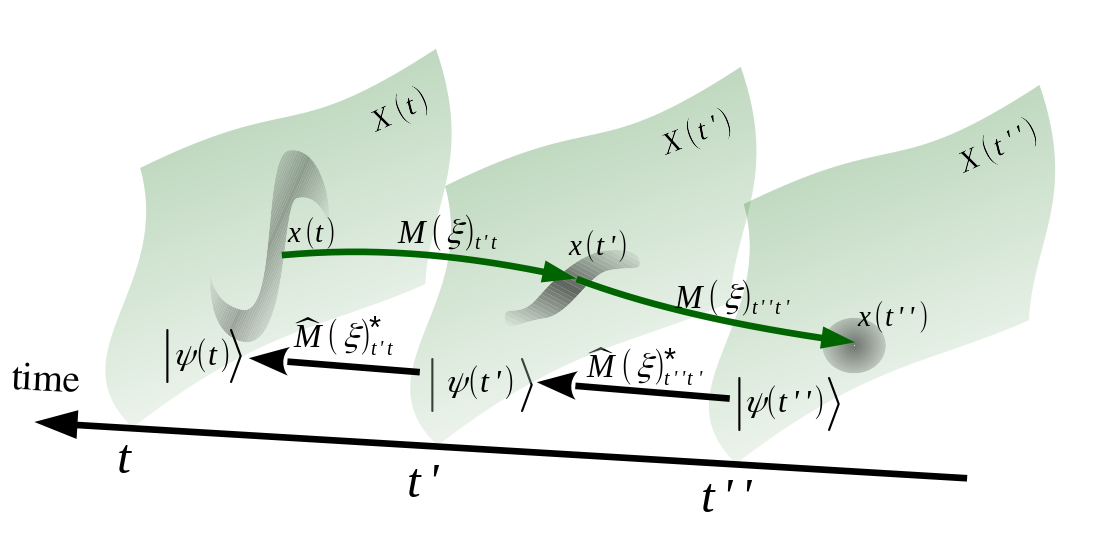}
    \caption{\label{Figure_1} (left) A continuous-time stochastic DS is defined by a flow vector field, $\mathscr F$, from the tangent space, $TX$, of the phase space, $X$. ${\mathscr F}(\xi(t))$ is time-dependent due to the presence of the time-dependent noise, $\xi(t)$. The DS is equivalent to two-parameter family of noise-configuration-dependent diffeomorphisms, $M(\xi)_{tt'}$, such that the trajectories are given by $x(t)=M(\xi)_{tt'}(x(t'))$.  (right) In the spirit of the pathintegral representation of temporal evolution, there is a copy of $X$ at every time moment and evolution is defined by pullbacks, $\hat M(\xi)_{t't}^*$, induced by inverse maps, $M(\xi)_{t't}$. The pullbacks act on a time-dependent differential form, $|\psi(t)\rangle $, understood as a "wavefunction" - a time-dependent object encoding information of the system's past. When averaged over noise configurations, a pullback yields the generalized transfer operator, which is unique and corresponds to the Stratonovich interpretation of stochastic dynamics.}
\end{figure}

An external observer does not know which noise configuration is realized in a given experiment. Consequently, a probabilistic framework is necessary -- one that incorporates the observer's uncertainty about the noise and, consequently, the DS. However, the noise itself is not uncertain: in any given experiment, the noise is just a deterministic function of time. Therefore, before introducing the observer’s uncertainty into the model -- a step we will take later -- Eq.(\ref{SDE}) is an ODE governed by a time-dependent flow vector field, $\mathscr F$ (see Fig.\ref{Figure_1}). \footnote{This picture differs from the traditional understanding of SDEs, which is examined in Sec. \ref{Sec:Pathintegral_Representation_of_Stochastic_Dynamics} and whose relation to Eq. (\ref{SDE}) is discussed in Sec. \ref{Sec:dilemma}. }

For any given initial condition and noise configuration, Eq.(\ref{SDE}) yields a unique solution. Moreover, even if the noise configuration is not differentiable with respect to time, the solution is differentiable with respect to the initial condition. \cite{Slavik} Therefore, there is a two-parameter family of $\xi$-dependent diffeomorphisms,
\begin{eqnarray}
M(\xi)_{tt'}:X\to X, M(\xi)_{tt'}\circ M(\xi)_{t't''}=M(\xi)_{tt''}, \left.M(\xi)_{ t t'}\right|_{t=t'} = \text{Id}_X,
\end{eqnarray}
such that $x(t) = M(\xi)_{tt'}(x')$ is the solution with the initial condition $x(t')=x'$.

\subsection{Generalized probability distributions}
The dynamics can now be defined as follows: if at time $t'$, the system is described by the probability distribution $P(x)$, then the average value of some function $f:X\to\mathbb{R}$ at a later time $t$ is
\begin{eqnarray}
&\bar f(t) = \int_X f\left(M(\xi)_{tt'}(x)\right) P(x) \dd^Dx= \int_X f(x) \hat M(\xi)_{t't}^*\left(P(x) \dd^Dx\right).
\end{eqnarray}
Here $\hat M(\xi)^*_{t't}$ is action or pullback induced by the ''inverse'' map, $M(\xi)_{tt'}^{-1}=M(\xi)_{t't}$, on the probability distribution understood in the coordinate-free setting as a top-degree  differential form, $P(x)\dd^Dx\equiv P(x) \dd x^1 \wedge ... \wedge \dd x^D$. In other words, instead of propagating the dynamical variables forward, one can equivalently propagate the probability distribution backward,
\begin{eqnarray}
&P(t) = \hat M(\xi)_{t't}^*P(t').
\end{eqnarray}
If the observer is interested only in the original dynamical variables, the description of the DS in terms of $P(x)$ may suffice. However, to explore properties such as Lyapunov exponents, this description of the DS must be extended. Namely, one must introduce dynamical fields that represent "differentials" -- objects that belong to the tangent space of $X$ -- evolving in the same way as the differentials of the original dynamical variables. Moreover, there is also a requirement that the properties of the new fields must reflect the fact that propagating two parallel differentials is pointless as it does not yield any additional information about Lyapunov exponents, as compared to propagating just one differential. 

The fields that satisfy these requirements are anticommuting differentials or differential volume elements in the definition of the differential forms or $k$-forms:
\begin{eqnarray}
    &\psi^{(k)}(x) = (1/k!)\psi^{(k)}_{i_1....i_k}(x)\dd x^{i_1}\wedge ... \wedge \dd x^{i_k}\in\Omega^{(k)}(x), \; 0\le k\le D,
\end{eqnarray}
where $\Omega^{(k)}(x)$ is the space of $k$-forms at point $x$. Now, the above-mentioned generalization of the dynamical probability distributions is,
\begin{eqnarray}
    &\psi = \sum\nolimits_{k=0}^D\psi^{(k)} \in  \Omega = \bigoplus\nolimits_{k=0}^D\Omega^{(k)}.\label{wavefunction}
\end{eqnarray}
Its temporal evolution follows from the above example and is given by the pullback of inverse maps,
\begin{eqnarray}
\psi(t) =   \hat M(\xi)_{t't}^* \psi(t').
\end{eqnarray}
Only top differential forms from $\Omega^{(D)}$ represent probability distributions. \footnote{In some cases, it may be possible to interpret the wavefunctions as the conditional probability distributions. \cite{OvcEntropy}} If at some moment of time they are positive everywhere on $X$, they will remain this property at all later times (see Sec.\ref{ErgodicZero} below). Other generalized distributions can be negative and the term "distribution" may be misleading. Therefore, we adopt the terminology of quantum theory and refer to them instead as "wavefunctions". As compared to the conventional probability distributions, the wavefunctions contain additional memory that encodes information about Lyapunov exponents. \cite{Lyapunov_SUSY} 

\subsection{Generalized transfer operator}
Unlike trajectories, points in $X$, or maps, pullbacks are linear objects even when $X$ is nonlinear. As a linear object, $\hat M(\xi)^*_{t't}$ can be averaged over noise configurations, yielding an evolution operator that incorporates the uncertainty of the external observer about the DS, \footnote{The order of time arguments is purposely reversed here so that (for a white noise): $\hat{\mathcal M }_{tt''}=\hat{\mathcal M }_{tt'}\hat{\mathcal M }_{t't''}$.}
\begin{eqnarray}
    \hat{\mathcal M }_{tt'} = \iint \hat M(\xi)^*_{t't} {\mathcal P}(\xi) \mathcal{D}\xi \overset{\text{def}}{=} \langle\hat M(\xi)^*_{t't}\rangle_{\text{noise}}.\label{generalGTO}
\end{eqnarray}
Here, ${\mathcal D}\xi$ and ${\mathcal P}(\xi)$ are, respectively, the differential of the functional integration (see, \emph{e.g.}, Ref.\cite{FunctionalIntegral} and Refs. therein) over the noise configurations and the corresponding normalized probability functional, so that $\langle 1\rangle_\text{noise}=1$. The properties of the noise can be defined either by specifying $\mathcal P$, as will be done in Sec.\ref{Sec:Pathintegral_Representation_of_Stochastic_Dynamics}, or by specifying all the noise averages, for example, via the introduction of the generating functional, $G(\eta) = \langle e^{ \int \eta_a(\tau) \xi^a(\tau) \dd\tau  }\rangle_\text{noise}$, so that,
\begin{eqnarray}
   & \langle \xi^{a_1}(t_1)...\xi^{a_k}(t_k)\rangle_{\text{noise}} = \frac{\delta}{\delta \eta_{a_1}(t_1)}...\frac{\delta}{\delta \eta_{a_k}(t_k)} G(\eta)\big|_{\eta=0},
\end{eqnarray}
where $\delta/\delta\eta_a(t)$ denotes functional differentiation. Below, we use the Gaussian white noise with $G(\eta) = e^{\int (\eta^2(\tau)/2) \dd\tau }$ and,
\begin{eqnarray}
&\langle\xi^a(t)\rangle_{\text{noise}}  = 0, \; \langle\xi^a(t)\xi^b(t')\rangle_{\text{noise}} = \delta^{ab}\delta(t-t'), ... \label{GussianWhiteNoise}
\end{eqnarray}
where $\delta(t-t')$ is the Dirac delta function.

In DS theory, the pullback averaged over noise is known as the generalized transfer operator (GTO). \cite{Rue02,Rue1990} We use the same identifier for operator (\ref{generalGTO}), which is a variant of this concept with the difference that the noise is infinite dimensional and the pullbacks correspond to inverse diffeomorphisms.

The explicit form of the GTO can be derived by utilizing the concept of the chronological ordering of operators,
\begin{eqnarray}
&\hat M(\xi)^*_{t't} \equiv {\mathcal T} e^{-\int_{t'}^t \hat L_{{\mathscr F}(\xi(\tau))}\dd\tau} \label{ChronologicalOrder}\\ \nonumber 
&= \hat 1_{\Omega} - \int_{t'}^t \hat L_{{\mathscr F}(\xi(\tau))}\dd\tau + \int_{t'}^t \hat L_{{\mathscr F}(\xi(\tau_1))}  \dd\tau_1 \int_{t'}^{\tau_1} \hat L_{{\mathscr F}(\xi(\tau_2))} \dd\tau_2 \dots,    
\end{eqnarray}
which is the solution of the following differential equation,
\begin{eqnarray}
&\partial_t \hat M(\xi)^*_{t't} = - \hat L_{{\mathscr F}(\xi(t))}\hat M(\xi)^*_{t't},\; \hat M(\xi)^*_{t't}\big|_{t=t'} = \hat 1_{\Omega}, \label{DiffEqPullback}
\end{eqnarray}
where $\hat L_{{\mathscr F}(\xi(t))}=\dd \hat M(\xi)^*_{tt'}/\dd t|_{t=t'}$ is the infinitesimal pullback or Lie derivative. Eq.(\ref{DiffEqPullback}) can be obtained by Taylor expanding the following equation in $\Delta \tau$
\begin{eqnarray}
    &\hat M(\xi)^*_{t't+\Delta \tau} = \widehat {(M(\xi)_{t't}\circ M(\xi)_{tt+\Delta \tau})^*} = \hat M(\xi)_{tt+\Delta \tau} ^* \hat M(\xi)_{t't}^*,
\end{eqnarray}
and taking the limit $\Delta \tau\to 0$ using $\hat M(\xi)_{tt+\Delta \tau}^* = \hat 1_{\Omega} - \Delta \tau \hat L_{{\mathscr F}(\xi(t))} +... $. The minus sign here reminds once again that we are dealing with the inverse diffeomorphisms.

Assuming Gaussian white noise (\ref{GussianWhiteNoise}), utilizing the linearity of the Lie derivative in its argument, $\hat L_{{\mathscr F}(\xi(t))}=\hat L_F + \xi^a(t)(2\Theta)^{1/2} \hat L_{G_a}$, noting that $\hat{\mathcal M }_{t t''} = \hat{\mathcal M }_{t t'} \hat{\mathcal M }_{t' t''}$ because the variables of white noise at different time moments do not correlate, and using Eq.(\ref{ChronologicalOrder}) and identities $\int_{t}^{t+\Delta \tau} \dd\tau_1 \int_{t}^{\tau_1}\dd\tau_2 =\Delta \tau^2/2$ and $\int_{t}^{t+\Delta \tau} \dd\tau_1 \int_{t}^{\tau_1} \delta(\tau_1-\tau_2)\dd\tau_2 = \Delta \tau/2$, one has,
\begin{eqnarray}
&\hat{\mathcal M }_{t+\Delta \tau,t'} = \hat{\mathcal M }_{t+\Delta \tau,t} \hat{\mathcal M }_{tt'} =\langle{\mathcal T} e^{-\int_{t}^{t+\Delta \tau} \hat L_{{\mathscr F}(\xi(\tau))}\dd\tau}\rangle_\text{noise} \hat{\mathcal M }_{tt'} \label{GTO_deriv_1}\\
&= \big(\hat 1_{\Omega} - \Delta \tau\hat L_F +\Delta \tau^2\hat L_F^2/2 + \Delta \tau \Theta \hat L_{G_a}\hat L_{G_a} + ...\big) \hat{\mathcal M }_{tt'}.\nonumber    
\end{eqnarray}
In the limit $\Delta \tau\to 0$, the above equation gives,
\begin{eqnarray}
&\partial_t \hat{\mathcal M }_{tt'} = - \hat H \hat{\mathcal M }_{tt'},\nonumber    
\end{eqnarray}
which integrates into the following expression for the \emph{finite-time} GTO, 
\begin{eqnarray}
&\hat{\mathcal M }_{tt'} = e^{-(t-t')\hat H}.\nonumber    
\end{eqnarray}
Here, the \emph{infinitesimal} GTO,
\begin{eqnarray}
    \hat H = \hat L_F - \Theta \hat L_{G_a}\hat L_{G_a},\label{GTO_0}
\end{eqnarray}
has a very clear meaning: the first and the second terms are, respectively, the Lie derivative representing the drift along the deterministic part of the flow, $F$, and the Laplacian\footnote{To be more accurate, this is a member of the family of Laplacians.} representing the diffusion associated with the accumulation of uncertainty due to the influence from the noise.

From the point of view of the theory of SDEs, the GTO (\ref{GTO_0}) is a stochastic evolution operator in the Stratonovich interpretation of stochastic dynamics.\cite{elworthy1999geometry} However, unlike the stochastic evolution operators in the theory of SDEs and/or the Parisi-Sourlas approach, the GTO has a clear-cut mathematical meaning, making it unique and eliminating the need for additional interpretations beyond its definition. 

\subsection{Topological supersymmetry}
Central to our discussion is the Cartan formula for a Lie derivative, \emph{e.g.}, 
\begin{eqnarray}
\hat L_{F} = [\hat d, \hat{\imath}_{F}],\label{Cartan}
\end{eqnarray}
where $\hat d: \Omega^{(k)}\to\Omega^{(k+1)}, k=0,D-1$ and $\hat{\imath}_{F}: \Omega^{(k)}\to\Omega^{(k-1)}, k=1...D$ are, respectively, the exterior derivative and interior multiplication along its argument (see, e.g., Ref.\cite{Nakahara}), and we introduced the concept of bi-graded commutator:
\begin{eqnarray}
    [\hat a, \hat b] = \hat a \hat b - (-1)^{\text{deg }{\hat a}\text{ deg }{\hat b}} \hat b \hat a,
\end{eqnarray}
where the degree of an operator is defined as $\text{deg } \hat a = l-k$ for $\hat a: \Omega^{(k)}\to \Omega^{(l)}$. Both, $\hat d$ and $\hat{\imath}_{\mathscr F}$ have odd degrees so that the r.h.s. of Eq.(\ref{Cartan}) is an anti-commutator.

The fundamental property of $\hat d$ is nilpotency, $\hat d^2=0$. It implies, particularly, that $\hat d$ commutes with any $\hat d$-exact operator, i.e., operator of the form $[\hat d, \hat a]$: $[\hat d,[\hat d, \hat a]]=0, \forall \hat a$. This property and the fact that a commutator with $\hat d$ is a bi-graded "differentiation", $[\hat d, \hat a\hat b] = [\hat d, \hat a]\hat b + (-1)^{\text{ind} \hat a} \hat a[\hat d, \hat b]$, can be used to reveal that the GTO in Eq.(\ref{GTO_0}) is $\hat d$-exact, 
\begin{eqnarray}
    \hat H = [\hat d, \hat {\bar d}],\label{GTO}
\end{eqnarray}
with $\hat {\bar d} = \hat{\imath}_{F} - \Theta \hat{\imath}_{G_a}\hat L_{G_a}$. As a $\hat d$-exact operator, the GTO (\ref{GTO}) is also $\hat d$-closed, i.e., it is commutative with $\hat d$, \footnote{Getting a bit head, the $\hat d$-exactness of the GTO (\ref{GTO}) is a stronger property than its $\hat d$-closeness (\ref{Commutativity}). While $\hat d$-closeness ensures the pairing of the eigenstates into non-supersymmetric doublets (see Sec. \ref{Sec:susyDoublets} below), $\hat d$-exactness further implies that supersymmetric singlets have exactly zero eigenvalues (see Sec.\ref{Sec:susySinglets} below).}
\begin{eqnarray}
    [\hat d, \hat H] = 0.\label{Commutativity}
\end{eqnarray}
In physics, it is said that a model has an internal continuous symmetry if there is a continuous group of operators such that $\hat G \hat H \hat G^{-1}=\hat H$, where $\hat G$ represents an element of the group. The generators of this group commute with $\hat H$ whose eigenstates form irreducible representations of the group: one-dimensional fully symmetric representations corresponding to non-degenerate eigenstates are called singlets, while more-than-one dimensional representations corresponding to degenerate eigenstates are called multiplets. The symmetry is said to be broken spontaneously if the ground state is degenerate, i.e., it is a multiplet.

This scenario applies to our case: the continuous group of internal symmetry is $\hat G_s=e^{s\hat d} = 1 + s \hat d$, where $s\in \mathbb{R}$, $\hat G_s \hat G_t = \hat G_{t+s}$, and $\hat G_{-s} \hat H \hat G_{s} = \hat H$; the sole generator of this group is $\hat d$; the supersymmetric singlets in Sec.\ref{Sec:susySinglets} are the fully symmetric one-dimensional representations, while the non-supersymmetric doublets in Sec.\ref{Sec:susyDoublets} are irreducible 2-dimensional representations. Consequently, $\hat d$ can be recognized as the (generator of the) symmetry of the model. 

This symmetry can be further identified as a supersymmetry -- the term used for symmetries that mix bosons and fermions. This becomes evident when the differentials of wavefunctions are represented by Grassmann numbers or fermions, \cite{Wit82} $\wedge \dd x^i\sim\chi^i$, so that the basic property of the differentials, $\dd x^{i_1} \wedge \dd x^{i_2} = -\dd x^{i_2} \wedge \dd x^{i_1}$, is consistent with that of Grassmann numbers, $\chi^{i_1}\chi^{i_2}=-\chi^{i_2}\chi^{i_1}$. In these terms, the exterior derivative has the form,
\begin{eqnarray}
&\hat d = \chi^i \partial/\partial x^i,
\end{eqnarray}
and it acts on a wavefunction by destroying a boson, $x$, and creating a fermion, $\chi$. 

Furthermore, being a fundamental object in algebraic topology -- where it serves as an algebraic dual of the boundary operation -- $\hat d$ can be identified as the topological supersymmetry (TS). Another justification for this terminology is that a $\hat d$-exact evolution operator, as in Eq.(\ref{GTO}), is a defining characteristic of TFTs, \cite{TFT_BOOK} where the pathintegral counterpart of $\hat d$ -- typically denoted as $Q$ (see Eq.(\ref{TS_pathintegral}) below) -- is referred to as TS.

From the mathematical point of view, the presence of TS follows from the fact that $\hat d$ commutes with pullbacks of all diffeomorphisms. In other words, the TS is an algebraic representation of the property of diffeomorphisms to preserve the topology of the phase space: infinitely close initial conditions result in close trajectories for any noise configuration. This further suggests that if TS is broken (see below), this property may no longer hold, allowing initially close points to evolve into trajectories that diverge over infinitely long evolution -- a hallmark of chaos known as the butterfly effect. \cite{Lorenz} Through its algebraic structure, STS extends this traditional understanding of the butterfly effect based on the concept of a deterministic trajectory to stochastic dynamics, where all trajectories are possible.

\subsection{Eigensystem}
\label{Sec:Eigensystem}
The eigensystem of the GTO has a set of properties that constrains the spectra of the physically meaningful models -- those with discrete spectra and finite spectral radius of the GTO -- to the three major types presented in Fig.\ref{Figure_2}. \cite{OvcEntropy} These properties are discussed in this section.

\subsubsection{Pseudo-Hermiticity and completeness} 
{ The GTO is a real operator and its eigenvalues are either real or come in complex conjugate pairs. This makes it a pseudo-Hermitian operator.} \cite{Mos02} {Complex conjugate pairs of eigenvalues can be identified as Reulle-Pollicott resonances and they can be thought of as a nontrivial representation of the pseudo-time reversal ($\eta T$-) symmetry.} \cite{Mos03} {This symmetry and its breaking will be recalled in} Sec.\ref{EtaTSymmetry} {below. For now, the only property of pseudo-Hermiticity of $\hat H$ we need is the existence of a complete bi-orthogonal eigensystem:}
\begin{eqnarray}
    &\hat H\psi_\alpha  = H_\alpha \psi_\alpha,  \bar \psi_\alpha \hat H = \bar \psi_\alpha H_\alpha, \; \int_X \bar \psi_\alpha \wedge \psi_\beta = \delta_{\alpha\beta},\label{eigensystem}
\end{eqnarray}
with $H_\alpha$ being the corresponding eigenvalues. For simplicity, the spectrum is assumed discrete, which is the case for compact $X$ and non-degenerate noise, that is, such that the noise-induced metric $g^{ij}(x) = G^i_a(x)G_a^j(x)$ is non-degenerate $\text{min} \text{ spec } g(x) > 0, \forall x$ so that $\hat H$ is elliptic.

Any (right) wavefunction, $\psi$, can be resolved as
\begin{eqnarray}
    &\psi = \sum_{\alpha} c_\alpha \psi_\alpha, c_\alpha = \int_X \bar\psi_\alpha \wedge \psi.\label{completeness}
\end{eqnarray}
To make the distinction between the left and right wavefunctions manifest, let us adopt the terminology of quantum theory and refer to the left and right vectors as bras and kets, $\psi_\alpha\to|\alpha\rangle$, $\bar\psi_\alpha\to\langle \alpha|$. In these notations, Eqs.(\ref{eigensystem}) read
\begin{eqnarray}
    &\hat H|\alpha\rangle = H_\alpha |\alpha\rangle, \langle\alpha| \hat H= \langle \alpha| H_\alpha, \; \langle\alpha|\beta\rangle \overset{\text{def}}{=} \int_X \bar\psi_\alpha \wedge \psi_\beta = \delta_{\alpha\beta},
\end{eqnarray}
and the completeness property (\ref{completeness}) can be expressed as a resolution of unity,
\begin{eqnarray}
    &\hat 1_{\Omega} = \sum\nolimits_{\alpha} | \alpha\rangle \langle \alpha |, \; |\psi \rangle = \hat 1_{\Omega}  |\psi \rangle = \sum_{\alpha} c_\alpha |\alpha\rangle, c_\alpha = \langle\alpha|\psi\rangle.
\end{eqnarray}
\subsubsection{Conservation of fermions} 
The GTO is block diagonal, $\hat H=\text{diag}(\hat H^{(0)}...\hat H^{(D)})$, because it commutes with the operator of the degree of a differential form, $\hat k |\psi\rangle = k |\psi\rangle, \forall |\psi\rangle\in \Omega^{(k)}$. Each eigenstate has a well-defined degree,
\begin{eqnarray}
    &\hat k | \alpha \rangle  = k_\alpha | \alpha \rangle,\label{degree_of_eigenstate} \;  0\le k_\alpha\le D.
\end{eqnarray}
In physics terms, $\hat k$ is the operator of the number of fermions, which is a conserved quantity due to its commutativity with the GTO.
\subsubsection{Non-supersymmetric doublets} \label{Sec:susyDoublets}
Except for a few supersymmetric singlets (see below), all the eigenstates are non-supersymmetric "doublets". {This can be seen as follows.} \cite{Torsten} {Let first note that if $|\alpha\rangle$ is an eigenstate and $|\alpha'\rangle = \hat d|\alpha\rangle \ne 0$ than $|\alpha'\rangle$ is also an eigenstate with the same eigenvalue because $[\hat d, \hat H]=0$ and, }
\begin{eqnarray}
    & \hat H |\alpha\rangle = H_\alpha | \alpha\rangle \to \hat d \hat H |\alpha\rangle = H_\alpha \hat d | \alpha\rangle \to  \hat H |\alpha'\rangle = H_\alpha |\alpha'\rangle.
\end{eqnarray}
We also note that $\hat d$ raises the degree of a wavefunction: $k_{\alpha'}=k_{\alpha}+1$.

\begin{figure}[t] 
    \centering
    \includegraphics[width=0.45\linewidth]{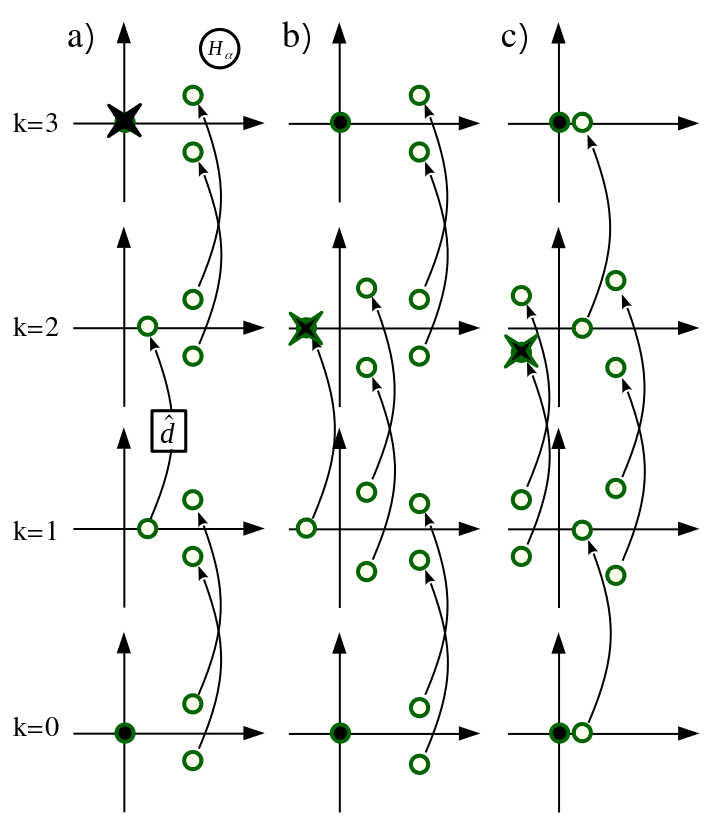}
    \caption{\label{Figure_2} The three possible types of GTO spectra (a,b,c) for a stochastic DS with $X=\mathbb{S}^3$. Each row ($k = 0 , . . .3$) contains three graphs representing $\text{spec}H^{(k)}$ for the three different types of spectra. Dots at the origin for $k=0$ and $k=3$ indicate the supersymmetric eigenstates from the zeroth and the third de Rham cohomologies of $X$. In cases b and c, the ground states (crosses) are non-supersymmetric doublets, as they possess non-zero eigenvalues, signifying spontaneous breakdown of TS. Additionally, in case c, the pseudo-time reversal symmetry is also broken. The vertical arrowed curves illustrate the action of the TS operator.}
\end{figure}

Let us now act by $\hat d$ on each eigenstate of $\hat H^{(0)}$, as visualized by the lowest row of curved arrows in Fig.\ref{Figure_2}. The result is a set of all the $\hat d$-exact eigenstates of $\hat H^{(1)}$. \footnote{A $\hat d$-exact state is a state from the image of $\hat d$, i.e., a state of the form $\hat d|a\rangle$.} We can now further act by $\hat d$ on all the eigenstates of $\hat H^{(1)}$, as visualized by the second row of curved arrows in Fig.\ref{Figure_2}. All the $\hat d$-exact eigenstates which came from $\hat H^{(0)}$ vanish due to the nilpotency of TS: $\hat d^2=0$. Other eigenstates turn into $\hat d$-exact eigenstates of $H^{(2)}$. Continuing this recursive procedure, we traverse and pair up all the eigenstates except those which are $\hat d$-closed \footnote{A $\hat d$-closed state is a state in the kernel of $\hat d$, i.e., a state, $|a\rangle$, such that $\hat d |a\rangle = 0$.}  but not $\hat d$-exact, i.e., the ones that are nontrivial in the cohomology of $\hat d$, or, in de Rham cohomology. These are the supersymmetric singlets which we will speak about shortly.

This procedure can be repeated for bras. It goes in reverse direction, however, and one starts with the bras of the eigenstates for $\hat H^{(D)}$, which are degree zero differential forms. \footnote{Dropping the bra-ket notations, if the ket of an eigenstate $\psi_\alpha\in\Omega^{(k)}$, then its bra $\bar \psi_\alpha\in\Omega^{(D-k)}$.} Subsequently acting by $\hat d$ from the right, we obtain the pairs of bras such as $\langle \alpha' |, \langle \alpha | = \langle \alpha' |\hat d$. They are the bras of the corresponding kets, and their normalization takes the form $\langle\alpha'|\alpha'\rangle = \langle\alpha|\alpha\rangle = \langle\alpha'|\hat d|\alpha\rangle = 1 $. In this manner, each doublet can be defined via a single bra-ket pair, $\langle \tilde \alpha |$ and $|\tilde \alpha \rangle$, such that
\begin{eqnarray}
    & |\alpha \rangle = |\tilde \alpha \rangle, \langle \alpha | = \langle \tilde \alpha |\hat d, \text { and } \; |\alpha' \rangle = \hat d |\tilde \alpha \rangle, \langle \alpha' | = \langle \tilde \alpha |,
\end{eqnarray}
and the orthogonality property reads,
\begin{eqnarray}
    & \langle \tilde \alpha| \tilde \beta\rangle = 0 , \langle \tilde \alpha| \hat d | \tilde \beta\rangle = \delta_{\tilde \alpha \tilde \beta}.
\end{eqnarray}
\subsubsection{Supersymmetric singlets} 
\label{Sec:susySinglets}
As we already mentioned, the only eigenstates that are not paired up into non-supersymmetric doublets by the above procedure are those from de Rham cohomology. Due to the completeness of the eigensystem, each de Rham cohomology class contributes one supersymmetric "singlet." The bra and ket of each such eigenstate satisfy $\hat d |\theta\rangle = 0, \langle \theta | \hat d = 0$. This means, particularly, that the expectation value of any $\hat d$-exact operator vanishes, $\langle \theta | [\hat d, \hat a]| \theta\rangle =0, \forall \hat a$. Since $\hat H$ is $\hat d$-exact (see Eq.\ref{GTO}), all supersymmetric singlets have zero eigenvalue.

Summing up the properties of the eigensystem discussed so far, the resolution of unity on $\Omega$ can now be expressed as:
\begin{eqnarray}
    &\hat 1_{\Omega} = \sum\nolimits_{\theta} |\theta\rangle \langle\theta | + \sum\nolimits_{\tilde \beta} \left(\hat d|\tilde \beta \rangle\langle \tilde \beta |+|\tilde \beta\rangle \langle\tilde \beta |\hat d\right) \\ &\nonumber  + \sum\nolimits_{\tilde \gamma, \pm} \left(\hat d|\tilde \gamma, \pm\rangle \langle\tilde \gamma, \pm |+|\tilde \gamma, \pm\rangle \langle\tilde \gamma, \pm |\hat d\right),
\end{eqnarray}
where $\theta$, $\tilde \beta$, and $\tilde\gamma, \pm$ run over the supersymmetric singlets and non-supersymmetric doublets with the real and complex-conjugate eigenvalues, respectively. The operator $\hat {\bar d}$ from Eq.(\ref{GTO}) can be expressed, up to a $\hat d$-exact piece which does not change the GTO, $\hat H = [\hat d, \hat{\bar d} + [\hat d, \hat a] ] = [\hat d, \hat{\bar d}]$, as,
\begin{eqnarray}
    &\hat{\bar d} = \sum\nolimits_{\tilde \beta} |\tilde \beta\rangle H_{\tilde \beta} \langle\tilde \beta | + \sum\nolimits_{\tilde \gamma,\pm} |\tilde \gamma,\pm\rangle H_{\tilde \gamma,\pm} \langle\tilde \gamma,\pm |,\label{dbar_another_representation}
\end{eqnarray}
where $ H_{\tilde \beta} =  H_{\tilde \beta}^*, H_{\tilde \gamma, \pm} = H_{\tilde \gamma, \mp}^*$.

\subsubsection{Ergodic zero} 
\label{ErgodicZero}
The zero eigenvalue supersymmetric eigenstate of $\hat H^{(D)}$, the existence of which is topologically protected by the requirement of the completeness of the eigensystem of GTO, is always the "ground state" of $\hat H^{(D)}$,
\begin{eqnarray}
&\text{min Re} (\operatorname{spec} \hat H^{(D)}) = 0.\label{ErgZerocondition}
\end{eqnarray}
Therefore, any probability distribution evolves into this steady-state probability distribution, known also as "erdogic zero" or invariant measure. \cite{CarrascoRodriguezHertz2021} 

To prove this statement, we first observe that a probability distribution, with the property of being positive everywhere on $X$, will retain this property throughout its evolution. Indeed, this property is preserved by the pullback induced by any diffeomorphism: the pullback involves the transformation of variables within a positive function, followed by multiplication by the Jacobian of the variable transformation, which is also positive as all diffeomorphism preserve orientation. Consequently, this property is also preserved by a pullback averaged over noise, i.e., by the GTO.

Lets assume that, in contradiction with Eq.(\ref{ErgZerocondition}), 
\begin{eqnarray}
    &\Delta^{(D)}=-\text{min Re} (\operatorname{spec} \hat H^{(D)}) > 0.\label{Assumption}
\end{eqnarray}
Then, there is either a pair of eigenstates with complex conjugate eigenvalues such that $\text{Re }H_\alpha=-\Delta^{(D)}$ or a single eigenstate with a real eigenvalue, $H_\alpha=-\Delta^{(D)}$.

In the first case, an everywhere-positive probability distribution, $P\in\Omega^{(D)}$, will eventually begin to oscillate at sufficiently large times when its temporal evolution,
\begin{eqnarray}
    &P(t) = \sum_{\alpha, k_\alpha=D} e^{-tH_\alpha} c_\alpha \psi_\alpha, \; c_\alpha=\int_X \bar\psi_\alpha \wedge P(0),\label{ProbDistribution}
\end{eqnarray}
is dominated by the fastest growing pair of eigenstates with complex conjugate eigenvalues. This contradicts that $P(t)$ must be everywhere positive on $X$. 

In the second case, lets note that all non-supersymmetric eigenstates with non-zero eigenvalue from $\Omega^{(D)}$ are $\hat d$-exact. Indeed, $\hat H \psi_\alpha = [\hat d, \hat{\bar d}] \psi_\alpha = \hat d \hat{\bar d}\psi_\alpha = H_{\alpha}\psi_\alpha$, where we used that $\hat d\psi_\alpha=0$ because $\psi_\alpha\in\Omega^{(D)}$. Therefore, $\psi_\alpha = \hat d \tilde \psi_\alpha$ with $\tilde \psi = \hat{\bar d}\psi_\alpha /H_\alpha$. This further implies that $\int_X \psi_\alpha = \int_X \hat d \tilde \psi_\alpha = 0$. Consequently, all non-supersymmetric eigenstates are negative somewhere on $X$. In result, Eq.(\ref{ProbDistribution}) will take on negative values somewhere on $X$ at sufficiently large times when the contribution from the fastest growing non-supersymmetric eigenstate dominates $P(t)$. Therefore, Eq.(\ref{Assumption}) is not realizable, which proves Eq.(\ref{ErgZerocondition}). 

\subsubsection{Stochastic Poincare-Bendixson theorem} 
\label{Poincare_Bendixson:Theorem}
The property (\ref{ErgZerocondition}) also holds for $\hat H^{(0)}$ which is isospectral with $\hat H^{(D)}$ (see Ref.\cite{OvcEntropy}). This leads to conclusion that the spontaneous TS breaking is not possible for models with $\dim X<3$: there is simply no room for a non-supersymmetric pair of eigenstates with degrees differing by unity and with a real part of their eigenvalue being negative unless the dimensionality of the phase space is 3 or higher. This statement can be looked upon as a STS proof of the stochastic Poincare-Bendixson theorem. \cite{Stochastic_PB_theorem} 

{This may be a good moment to comment on the applicability of STS to random discrete-time dynamical systems. For such systems defined by maps that are diffeomorphisms, most conclusions drawn for SDEs remain directly applicable. If, however, the maps are not diffeomorphisms, a qualitatively new situation may arise.}\cite{Max_2019} {In particular, the TS symmetry operator may fail to commute with the evolution operator from the outset. This corresponds to what is known in theoretical physics as explicit symmetry breaking. All textbook examples of chaos with dimensionality lower than that allowed by the Poincaré–Bendixson theorem ($\dim X < 3$) fall into this category. Explicit symmetry breaking is qualitatively different from spontaneous symmetry breaking: in particular, the Goldstone theorem does not apply to explicit symmetry breaking. To the best of the present author’s knowledge, STS is the only theoretical framework that provides such a qualitative distinction between discrete-time and continuous-time chaos. Perhaps, this distinction merits further investigation which, however, goes beyond the scope of this paper.}

\subsection{Stochastic chaos}
From the DS theory viewpoint, dynamics can be characterized as chaotic if the spectral radius of the finite-time GTO is larger than unity. Under this condition, the partition function,  
\begin{eqnarray}
&Z_{tt'} = \text{Tr }\hat{\mathcal M }_{tt'} = \sum\nolimits_{\alpha}e^{-(t-t')H_\alpha},
\end{eqnarray}
grows exponentially in the limit of infinitely long evolution signaling the exponential growth of the number of closed solutions -- the hallmark of chaotic dynamics. This condition can be expressed as, 
\begin{eqnarray}
&\Delta = - \min\nolimits_\alpha \text{Re }H_\alpha > 0,\label{positivePressure}
\end{eqnarray}
where $\Delta$ is the rate of the exponential growth called "pressure". \cite{Rue02} It can be viewed as a random DS theory version of the family of dynamical entropies including topological entropy \cite{Handbook_of_DS,book_hyperbolic_flows} related, via Pesin formula, \cite{Pesin} to (stochastic) Lyapunov exponents. \cite{Arn03} 

Spectra b and c in the Fig.\ref{Figure_2} satisfy condition (\ref{positivePressure}). A practical example of spectrum of type b is the geodesic flow on a compact manifold of variable negative curvature. \cite{Rue02,Anosov_geodesic_flows} An example of type c spectrum is the kinematic dynamo, where the galactic magnetic field not only grows but also rotates. \cite{Torsten}

When condition (\ref{positivePressure}) is satisfied, the contribution into the partition function from the "erdogic zero" in Sec.(\ref{ErgodicZero}) is negligible in the long-time limit because it has a zero eigenvalue. This means that this state cannot represent the DS in the long-time limit, implying that investigating "ergodic zero" may not be the best way to explore chaos. The main contribution actually comes from the eigenstates with nontrivial degrees, $k\ne0,D$, because of Eq.(\ref{ErgZerocondition}). Getting a bit ahead, this foreshadows the high-energy physics picture that the spontaneous TS breaking leads to the emergence of a Dirac/Fermi sea of fermions.

{It should also be noted that for compact $X$, the existence of zero-eigenvalue supersymmetric states is topologically protected—the Hilbert space would simply be incomplete in their absence. Consequently, situations in which the ‘pressure’ becomes negative cannot occur. For non-compact $X$, however, the situation may differ: the absence of zero-eigenvalue supersymmetric states could be associated with issues of normalizability. This question merits further investigation, which lies beyond the scope of the present work.}

While alternative definitions of stochastic chaos may exist, positive pressure offers a significant advantage. Within this definition, the ground state of the model, which is (one of) the fastest growing eigenstates of the GTO, has a nonzero eigenvalue and is therefore non-supersymmetric. By definition, this implies the spontaneous breakdown of TS. As a result, positive pressure makes a good physical sense and it has a potential, via the Goldstone theorem, to explain the experimental signature of chaotic behavior known as 1/f noise as discussed in Sec.\ref{Sec:Phase_Diagram} below.

\subsection{Sharp trace}
Another key quantity is the \emph{sharp} trace of the GTO,
\begin{eqnarray}
&W = Tr (-1)^{\hat k} \hat{\mathcal M }_{tt'} = \sum\nolimits_\alpha (-1)^{k_\alpha}e^{-(t-t')H_\alpha}, 
\end{eqnarray}
where $\hat k$ and $k_\alpha$ are defined in Eq.(\ref{degree_of_eigenstate}). This quantity is a fundamental object of topological nature known in physics as the Witten index. 

For a non-supersymmetric doublet, the degrees of the paired eigenstates differ by unity. As a result, their contributions cancel out, leaving only supersymmetric singlets to contribute to the sharp trace. This leads to the expression:
\begin{eqnarray}
&W=\sum\nolimits_{k=0}^D (-1)^k B_k=Eu.Ch(X),
\end{eqnarray}
where $Eu.Ch.$ denotes the Euler characteristic of $X$ and $B_k$ is the Betti number, which counts the number of supersymmetric singlets of degree $k$.

\begin{figure}
    \centering
    \includegraphics[width=0.65\linewidth,height=4cm]{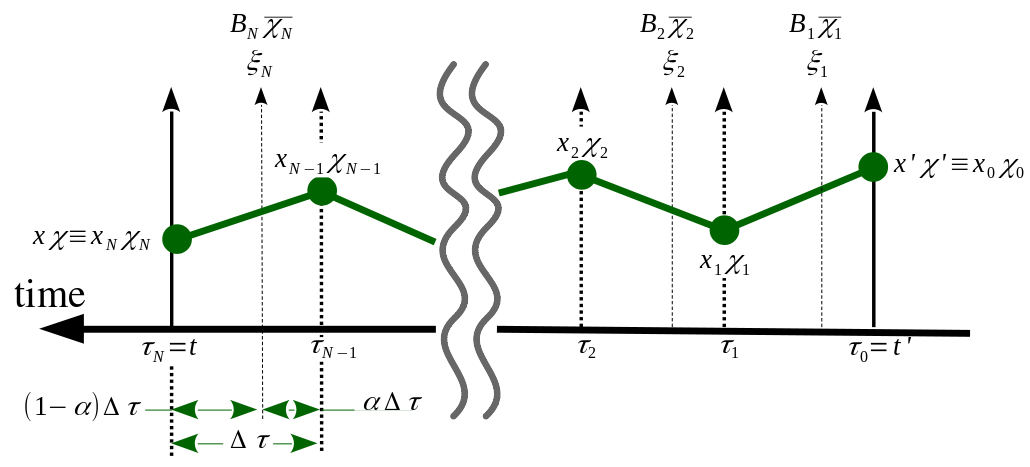}
    \caption{\label{Figure_3} Pathintegral is a continuous-time limit, $N\to\infty, \Delta \tau = (t-t')/N\to0$, of the discrete time evolution picture: the domain of temporal evolution, $(t,t')$, is split into $N$ time steps and the time takes on discrete values $\tau_N, \tau_{N-1},...\tau_1,\tau_0$, $t=\tau_N, t'=\tau_0$. Each time slice hosts a supersymmetric pair of fields $x_k, \chi_k$, and each dual slice hosts the corresponding supersymmetric pair of momenta fields, $B_k, \bar\chi_k$, along with the noise variable, $\xi_k$. The finite-time stochastic evolution operator is derived by integrating out all the fields except those at the first and last slices. Its exact expression depends on parameter $\alpha\in(0,1)$, which dictates how $x$ and $\chi$ are interpreted at the dual slice, $\tau_k$: $\alpha x_{k} + (1-\alpha)x_{k-1}$. Only for $\alpha=1/2$, corresponding to the Stratonovich interpretation of SDEs, does the stochastic evolution operator matches the generalized transfer operator of the DS theory, thereby having a clear-cut mathematical meaning of the pullback averaged over noise.}
\end{figure}

\section{Pathintegral representation of stochastic dynamics}
\label{Sec:Pathintegral_Representation_of_Stochastic_Dynamics}
\subsection{Stochastic differential equations}
In terms of the traditional theory of SDEs, Eq.(\ref{SDE}) with the Gaussian white noise can be expressed as,
\begin{eqnarray}
    \dd x(t) = F(x(t)) \dd t + (2\Theta) G_a(x(t)) \circ \dd W^a(t),\label{Traditional_SDE} 
\end{eqnarray}
where $W^a(t)$ is the Wiener process, a function whose derivative over time is the Gaussian white noise, and symbol $\circ$ indicates that this is a Stratonovich SDE. (see, e.g., Ref.\cite{Oks10} and Refs. therein) While notations in Eq. (\ref{Traditional_SDE}) may seem like time is continuous, there is an important subtlety: in the traditional theory of SDEs, stochastic dynamics is understood as a continuous-time limit of a discrete-time evolution, and this limit is taken only after averaging over the noise configurations (see Sec.\ref{Sec:dilemma} below). 

The discrete-time evolution picture of stochastic dynamics, assumed in Eq.(\ref{Traditional_SDE}), naturally aligns with numerical implementations of SDEs, as it formalizes a Runge-Kutta propagation scheme (augmented by stochastic noise). Furthermore, it is the basis for the pathintegral representation of stochastic dynamics, obtained through functional integration over trajectories. This formulation is the central focus of this section.

The discrete-time picture of evolution (see Fig. \ref{Figure_3}) is constructed by dividing the temporal domain into a large number, $N$, of time steps, forming a discrete lattice of time points, $\tau_k=\tau_0 + k \Delta \tau$, where $\Delta \tau = (t-t')/N$ is assumed to be vanishingly small but still finite. The discrete-time counterpart of Eq.(\ref{SDE}) is, \footnote{Unlike Eq.(\ref{SDE}), Eq.(\ref{SDE_discrete_time}) may look suspicious for nonlinear $X$'s because one cannot subtract points in a nonlinear space. The way around this subtlety is to believe that $x$'s are not the points themselves but are their coordinates within some coordinate neighborhood. This may raise concerns about potential loss of coordinate independence of the so-obtained description. In the continuous-time limit, however, the coordinate independence is restored (see Eq.(\ref{SEO_shifted}) below).}
\begin{eqnarray}
x_k - x_{k-1} = {\mathscr F}(\tilde x_k, \xi_k) \Delta \tau, \label{SDE_discrete_time}   
\end{eqnarray}
where the discrete-time version of the Gaussian white noise can be defined by the following probability distribution, 
\begin{eqnarray}
    &P(\xi) = (c(\Delta \tau))^N e^{-\sum_{k=1}^N \Delta \tau (\xi_k)^2/2},
\end{eqnarray}
with $c(\Delta \tau)=(\Delta \tau/2\pi)^{D/2}$ being the normalization constant, such that,
\begin{eqnarray}
&\int \prod_{k=1}^NP(\xi) \dd^D\xi_k = 1,\\
& \int \prod_{k=1}^N\xi^a_l\xi^b_{m}P(\xi)\dd^D\xi_k =\delta_{lm}\delta^{ab}/\Delta \tau.
\end{eqnarray}
In Eq.(\ref{SDE_discrete_time}), $\tilde x_{k} = \alpha x_{k} + (1-\alpha)x_{k-1}$ is an $\alpha$-family of approximations for $x$ during the $k^{th}$ time step. Different choices of $\alpha$ correspond to different "interpretations" of SDEs, with $\alpha = 0, 1/2, 1$ representing the Ito, Stratonovich, and Kolmogorov interpretations, respectively. \cite{Oks10}

The functional integration over noise variables can be defined as, 
\begin{eqnarray}
    &\langle 1 \rangle_{\text{noise}}  = \iint {\mathcal P}(\xi) {\mathcal D}\xi  \overset{\text{def}}{=}  \lim_{N\to\infty}\prod_{k=1}^{N} \int  P(\xi) \dd^D\xi_k= 1,
\end{eqnarray}
where the probability functional ${\mathcal P}(\xi)=e^{-\int_{t'}^t \dd\tau \xi^2(\tau)/2}$ and the normalization factors $c(\Delta \tau)$ are absorbed into the functional differential, ${\mathcal D}\xi$, for convenience.

\subsection{Parisi-Sourlas approach as a BRST gauge-fixing}
\label{Sec:ParisiSourlas}
One interpretation of the Parisi–Sourlas approach to SDEs is that it rewrites the partition function of the noise in terms of the model’s dynamical variables using the BRST gauge-fixing procedure. \cite{TFT_BOOK} This approach begins with the formal introduction of the dynamical variables into the partition function of the noise as, 
\begin{eqnarray}
&\iint {\mathcal P}(\xi) {\mathcal D}\xi \to \iint_{p.b.c} {\mathcal P}(\xi) {\mathcal D}\xi {\mathcal D}x,\label{introduction_of_x}
\end{eqnarray}
where the functional integration goes over closed paths or paths with periodic boundary conditions (p.b.c), 
\begin{eqnarray}
&\iint_{p.b.c} {\mathcal D}x \overset{\text{def}}{=} \lim\nolimits_{N\to\infty}\prod\nolimits_{k=1}^{N} \int_{X} \dd^D x_k.
\end{eqnarray} 
and there is no need to integrate over $x_0$ because the p.b.c. assume $x_0=x_N$. \footnote{Rewriting the noise partition function in terms of dynamical variables can be viewed as a change of integration variables within the noise partition function. If this transformation is expected to yield a scalar quantity -- rather than an operator, as in Sec.\ref{Sec:SEO} below -- then the numbers of $\xi$'s and $x$'s must be the same. This is the reason for using the p.b.c.}

At this stage, the right-hand side of Eq.(\ref{introduction_of_x}), though not well-defined and technically infinite, can be interpreted as a redundant theory of the noise. Its "action" is independent of $x$. This independence can be viewed as a local symmetry with respect to continuous deformations of the paths. This symmetry can be gauge-fixed using the SDE as a gauge condition, \cite{Henneaux1992} leading to the following pathintegral representation of the Witten index:
\begin{eqnarray}
&W=\iint_{p.b.c} J(\xi) \bigg( \prod\nolimits_\tau \delta ^D \bigg(\Big(\dot x(\tau) - {\mathscr F}(x(\tau),\xi(\tau)   \Big)\dd\tau\bigg) \bigg){\mathcal P}(\xi) {\mathcal D}\xi {\mathcal D}x .\label{Witten_1}
\end{eqnarray}
Here, the $\delta$-functional is introduced to limit the functional integration only to solutions of the SDE:\footnote{In this section, one may assume $\tilde x_k=x_k$ since the dependence on $\alpha$ disappears in the pathintegral representation of stochastic dynamics. It will reemerge later, when we transition further into the operator representation in Sec.\ref{Sec:SEO}.}
\begin{eqnarray}\label{functional_deltas}
&&\lim_{N\to\infty}\prod\nolimits_{k=1}^{N} \delta^D(M(\xi_k)_{t_{k-1} t_{k}}(x_{k})  - x_{k-1})\\
&\nonumber =&  \lim_{N\to\infty}\prod\nolimits_{k=1}^{N} \delta^D\bigg(\Big((x_{k}  - x_{k-1})/\Delta \tau - {\mathscr F}(x_k, \xi_k)\Big)\Delta \tau\bigg)\\
&\nonumber \overset{\text{def}}{=}& \prod\nolimits_\tau \delta ^D \bigg(\Big(\dot x(\tau) - {\mathscr F}(x(\tau),\xi(\tau))  \Big)\dd\tau \bigg),
\end{eqnarray}
where the single time-step map $M(\xi_k)_{t_{k-1} t_{k}}(x_{k}) = x_k - \Delta \tau {\mathscr F}(x_k, \xi_k) + ...$. Notation $J(\xi)$ in Eq.(\ref{Witten_1}) stands for the functional Jacobian, introduced to compensate, up to a sign, for the functional determinant that emerges when integrating out bosonic delta-functionals in (\ref{functional_deltas}) in a way which is a functional generalization of, $\int g(y) \delta^l(m(y))\dd^l y = \sum_{y_i,m(y_i)=0} g(y_i)/| J(y_i) |$, where $y\in \mathbb{R}^l$, $m:\mathbb{R}^l\to\mathbb{R}^l$, and $J(y)=\det_{(ij)} \partial m^i(y)/\partial y^j$ is the Jacobian of $m$. This functional Jacobian can be defined as,   
\begin{eqnarray}
J(\xi) &=& \lim_{N\to\infty} \det_{(k k')} \Big(\partial( M(\xi_k)_{t_{k-1} t_{k}}(x_{k})  - x_{k-1} )/\partial x_{k'} \Big) \\
\nonumber &=&\lim_{N\to\infty} \prod_{k=1}^{N} \delta^D \bigg( \Big( \partial \big(M(\xi_k)_{t_{k-1} t_{k}}(x_{k})  - x_{k-1} \big)/\partial x_{k'} \Big) \chi_{k'} \bigg)\dd^D \chi_k\\
\nonumber &=& \lim_{N\to\infty} \prod_{k=1}^{N} \delta^D( TM(\xi_k)_{t_{k-1}t_{k}}(x_k)\chi_{k} - \chi_{k-1})\dd^D \chi_k\\
\nonumber &=& \lim_{N\to\infty} \prod_{k=1}^{N} \delta^D\Big(\Big( (\chi_{k} - \chi_{k-1})/\Delta \tau - T{\mathscr F}(x_k, \xi_k) \chi_{k}\Big)\Delta \tau\Big)\dd^D \chi_k\\
\nonumber &\overset{\text{def}}{=}&\prod_{\tau} \delta^D\bigg(\Big( \dot \chi(\tau) - T{\mathscr F}(x(\tau),\xi(\tau))\chi(\tau) \Big)\dd\tau\bigg){\mathcal D}\chi,
\end{eqnarray}
where $TM(x) = \partial M(x)/\partial x$ is the tangent map and $T{\mathscr F}(x) = \partial {\mathscr F}(x)/\partial x$. The additional field $\chi\in TX$ is a Grassmann variable known as the Faddeev-Popov ghost. \cite{Faddeev1967} It has been introduced to make use of one of the key properties of Grassmann numbers: $\int \delta(\hat A^1_{i_1} \chi^{i_1})...\delta(\hat A^l_{i_l} \chi^{i_l}) \dd^l\chi \overset{\text{def}}{=}\int \delta^l(\hat A \chi) \dd^l\chi = \det_{(ij)} A^i_j$, where $A$ is a square $l\times l$ matrix (see, e.g., Ref.\cite{DeWitt_1992}). Its functional differential is defined as ${\mathcal D}\chi = \lim_{N\to\infty} \prod_{k=1}^{N} \dd^D \chi_k$.

The next step is to introduce momenta fields: the bosonic, $B$, and fermionic, $\bar\chi$, both defined on the dual time slices (see Fig.\ref{Figure_3}) and belonging to the cotangent space of $X$. These additional fields are needed to exponentiate the $\delta$-functionals in the way, which is a functional generalization of $\delta(f(x)) = \int e^{iB f(x)} \dd B/2\pi$ and its fermionic counterpart, $\delta^l(\hat A \chi)=\int e^{\bar\chi_i A^i_j \chi^j } \dd^l\bar\chi$. With their help,
\begin{eqnarray}
& \label{functional_delta}\prod\nolimits_\tau \delta ^D \bigg(\Big(\dot x(\tau) - {\mathscr F}(x(\tau),\xi(\tau))  \Big)\dd\tau\bigg) = \iint e^{i\int_{t'}^t B(\dot x - {\mathscr F}(x,\xi))\dd\tau} {\mathcal D}B,
\end{eqnarray}
where the integration measure is defined as $ {\mathcal D}B \overset{\text{def}}{=}  \lim_{N\to\infty}\prod_{k=1}^N \dd^DB_{k}/(2\pi)^D$, and similarly,
\begin{eqnarray}
\label{Jacobian_functional}J(\xi)&=&\prod_{\tau} \delta^D\bigg(\Big( \dot \chi(\tau) - T{\mathscr F}(x(\tau),\xi(\tau))\chi(\tau) \Big)\dd\tau\bigg){\mathcal D}\chi \\ 
&\nonumber  =& \iint e^{-i\int_{t'}^t \bar\chi (\dot \chi - T{\mathscr F}(x,\xi) \chi ) \dd\tau}{\mathcal D}\bar\chi {\mathcal D}\chi,
\end{eqnarray}
with $ {\mathcal D}\bar\chi \overset{\text{def}}{=}  \lim_{N\to\infty}\prod_{k=1}^N \dd^D\bar\chi_{k}$.\footnote{Strictly speaking, the differential also has the factor $(-i)^{DN}$, but one can always think that $N=4k, k\in N$.}

An important observation is that the product of the bosonic $\delta$-functional (\ref{functional_delta}) and the functional Jacobian (\ref{Jacobian_functional}) can be given as:
\begin{eqnarray}
&&\label{BRST_essence}J(\xi) \prod\nolimits_\tau \delta^D \bigg(\Big(\dot x(\tau) - {\mathscr F}(x(\tau),\xi(\tau))  \Big)\dd\tau\bigg) \label{Q_exact}\\
&\nonumber=&\iint e^{i\int_{t'}^t  \big( B(\dot x - {\mathscr F}(x,\xi)  ) - \bar\chi (\dot \chi - T{\mathscr F}(x,\xi) \chi ) \big)\dd\tau} {\mathcal D}B {\mathcal D}\bar\chi {\mathcal D}\chi
\\&\nonumber=&\iint e^{(Q, \Psi(\Phi,\xi))} {\mathcal D}B {\mathcal D}\bar \chi {\mathcal D}\chi,
\end{eqnarray}
where 
\begin{eqnarray}
&\Psi(\xi,\Phi) = \int_{t'}^{t} i\bar\chi(\dot x(\tau) - \mathcal F(x(\tau),\xi(\tau)))\dd\tau,
\end{eqnarray}
is the so-called gauge-fermion, the notation $\Phi=xB\chi\bar\chi$ represents the collection of all fields, and 
\begin{eqnarray}
&Q = \int_{t'}^t \Big(\chi(\tau)\delta/\delta x(\tau) + B(\tau)\delta/\delta \bar\chi(\tau)\Big)\dd\tau,\label{TS_pathintegral}
\end{eqnarray}
is the operator of the BRST symmetry and/or the pathintegral version of TS. \cite{TFT_BOOK} Eq.(\ref{BRST_essence}) is the core of the BRST gauge fixing procedure, where the gauge-fixing factors -- the bosonic delta-functional and the corresponding Jacobian -- are represented by an additional $Q$-exact piece in the action. 

The final step in obtaining the pathintegral representation of the Witten index is to integrate out the noise using the identity, $\iint {\mathcal D}\xi e^{\int (-\xi^2/2 + a\xi) \dd\tau }= e^{\int d\tau a^2/2}$, which leads from Eqs.(\ref{Witten_1}) and (\ref{Q_exact}) to, 
\begin{eqnarray}
&W =\iint_{p.b.c.}e^{(Q,\Psi(\xi,\Phi)) } {\mathcal P}(\xi) {\mathcal D}\xi {\mathcal D}\Phi = \iint_{p.b.c.} e^{(Q,\Psi(\Phi))}{\mathcal D}\Phi,\label{Parisi_Sourlas_Pathintegral}
\end{eqnarray}
where the new gauge fermion $\Psi = \int_{t'}^t (i\bar\chi \dot x - \bar d) \dd\tau$, with $\bar d = i\bar\chi ( F - \Theta G_a L_{G_a})$ and $L_{G_a}=(Q,i\bar\chi G_a)$ being the pathintegral versions of operator $\hat{\bar d}$ in Eq.(\ref{GTO}) and the Cartan formula for the Lie derivative (\ref{Cartan}), which follows from the recognition of $i\bar\chi G_a$ and $Q$ as the pathintegral versions of the interior multiplication by $G_a$ and the exterior derivative, respectively.

\subsection{Stochastic evolution operator}
\label{Sec:SEO}
The temporal evolution in the model is defined via the stochastic evolution operator (SEO) -- the Parisi-Sourlas pathintegral with open boundary conditions:
\begin{eqnarray}
&\iint_{{x\chi(t')=x'\chi'} \atop {x\chi(t)=x\chi}} e^{\int_{t'}^t (iB\dot x + i\dot \chi {\bar \chi} - H(\Phi)) \dd\tau}{\mathcal D}\Phi = \langle x\chi| e^{-(t-t')\hat H}|x'\chi'\rangle.\label{One_Boundary_Cond}
\end{eqnarray}
Here, we used $(Q,\Psi(\Phi))=\int_{t'}^t(iB\dot x + i\dot \chi {\bar \chi} - H(\Phi)) \dd\tau $, with $H=(Q,\bar d(\Phi))$, no integration is assumed over the variables on the first and the last time slices (see Fig.\ref{Figure_3}), $\hat H$ is the SEO in the operator representation, and the basis of the operator representation, where $x$ and $\chi$ are diagonal, is defined as: $\hat x|x \chi\rangle=x|x \chi\rangle, \hat \chi |x \chi\rangle=\chi|x \chi\rangle$, 
$\langle x\chi|\hat x = \langle x\chi|x, \langle x\chi|\hat \chi= \langle x\chi|\chi$. This basis is complete, 
\begin{eqnarray}
    &\langle x\chi|x'\chi'\rangle = \delta^D(x-x')\delta^D(\chi-\chi'), \; \int |x \chi\rangle\langle x\chi| \dd^D x \dd^D\chi =\hat 1_{\Omega},
\end{eqnarray} 
and any wavefunction can be resolved as,
\begin{eqnarray}
    &|\psi\rangle = \int \psi(x\chi) |x\chi\rangle  \dd^D x \dd^D \chi\text{,   } \psi(x\chi) = \langle x\chi|\psi\rangle.
\end{eqnarray}

The explicit form of $\hat H$ can be derived by considering a single step evolution of a wavefunction in the discrete time picture,
\begin{eqnarray}\label{TediousFormula}
&&\langle x\chi|e^{-\Delta \tau \hat H}|\psi\rangle = \int \langle x\chi|e^{-\Delta \tau \hat H}|y\eta\rangle \langle y\eta |\psi\rangle  \dd^D y \dd^D\eta \\ 
&&=\nonumber \int \left(e^{iB(x-y) + i \bar\chi(\chi-\eta) - \Delta \tau H(B\bar\chi \tilde x \tilde \chi)} \frac{\dd^DB}{(2\pi)^D}\dd^D(i\bar\chi) \right) \langle y\eta|\psi\rangle \dd^D y \dd^D\eta   \\ 
&&=\nonumber \int e^{iB(x-y) + i \bar\chi(\chi-\eta)}\Big(1 - \Delta \tau H(B\bar\chi \tilde x \tilde \chi) + ...\Big) \langle y\eta|\psi\rangle   \frac{\dd^DB}{(2\pi)^D}\dd^D(i\bar\chi)  \dd^D y \dd^D\eta \\ 
&&\nonumber =  (1 - \Delta \tau \hat H + ...) \langle x\chi|\psi\rangle,
\end{eqnarray}
where $\tilde x = \alpha x + (1-\alpha) y$, $\tilde \chi = \alpha \chi + (1-\alpha) \eta$, and $\hat H = \left.H(\tilde x\tilde \chi B\bar\chi)\right|_{B,\bar\chi\to\hat B,{\hat {\bar\chi}}}$, with
\begin{eqnarray}
    &i\hat B=\partial/\partial x, i\hat{\bar\chi}=\partial/\partial\chi,\label{momenta}
\end{eqnarray}
being the momenta operators whose form follows from expressions like this one, \footnote{Here, we explicitly show the vector indices.}
\begin{eqnarray}
    & \int e^{iB(x-y)} iB_{j_1}...iB_{j_p} x^{k_1} ... x^{k_q}  y^{l_1} ... y^{l_p}\langle y \eta|\psi\rangle \frac{\dd^DB}{(2\pi)^D} \dd^D y\label{OperatorOrdering}\\ 
    \nonumber &= x^{k_1} ... x^{k_q} \frac{\partial}{\partial x^{j_1}} ... \frac{\partial}{\partial x^{j_p}}x^{l_1} ... x^{l_p} \langle x \eta |\psi\rangle,
\end{eqnarray}
and similar expression can be derived for fermionic fields. 

Besides proving Eq.(\ref{momenta}), Eq.(\ref{OperatorOrdering}) also shows how to order operators: in the operator representation the position and momentum operators do not commute and the order of operators matters. As can be seen from the second line of Eq.(\ref{OperatorOrdering}), the correct order is chronological: $B,y$ and $x$ represent, respectively, $B_{k}$, $x_{k-1}$, and $x_{k}$ at any given time slice, $k$, of Fig.\ref{Figure_3}, so that $B$ acts after $y$ but before $x$.

The exact form of SEO depends on $\alpha$ because, say, $B\tilde x \to \alpha \hat x \hat B + (1-\alpha) \hat B \hat x$ (and similarly for fermions). Dropping the details, which can be found in Ref.\cite{OvcEntropy}, the SEO has the same form as the GTO (\ref{GTO_0}),
\begin{eqnarray}
    \hat H = \hat L_{F_\alpha} - \Theta \hat L_{G_a}\hat L_{G_a},\label{SEO_shifted}
\end{eqnarray}
but with a shifted flow vector field, $F_\alpha = F - \Theta(2\alpha-1)(G_a\cdot\partial) G_a$.

\subsection{Ito-Stratonovich dilemma}
\label{Sec:dilemma}
The dependence of SEO (\ref{SEO_shifted}) on $\alpha$ is the essence of Ito-Stratonovich dilemma. The meaning of this ambiguity in the definition of stochastic dynamics can be qualitatively understood as follows.

The entire family of Runge-Kutta methods (see, e.g., Ref.\cite{Thijssen_2007}) is based on the understanding that, under general assumptions, for any given initial condition and a sufficiently smooth configuration of the noise, $\xi(t)\to\xi(\tau_k)\equiv\xi_k, k=1...N$, the contituous-time limit of Eq.(\ref{SDE_discrete_time}) exists and the solution converges to that of Eq.(\ref{SDE}). The parameter $\alpha$ controls how the error approaches zero: \emph{e.g.}, for the direct Euler method, where $\alpha=0$, and the midpoint method, with $\alpha=1/2$, the error $\sim \Delta \tau$ and $\Delta \tau^2$, respectively. Importantly, the solution itself is unique and independent of $\alpha$. Therefore, if we choose to take the continuous-time limit before averaging over the noise, Eq.(\ref{SDE_discrete_time}) transforms into Eq.(\ref{SDE}), eliminating any dependence on $\alpha$. Now, the analysis of Sec.\ref{Sec:Dynamical_Systems_Theory} applies, so that the temporal evolution of differential forms is governed by the GTO (\ref{GTO_0}), which has a very clear mathematical meaning and is independent of the parameter $\alpha$.

This point of view on stochastic dynamics -- the one employed in Sec.\ref{Sec:Dynamical_Systems_Theory} -- can be described as first taking the continuous-time limit and then averaging over the noise. The pathintegral representation, however, reverses the order of these operations as can be seen, particularly, from Eq.(\ref{TediousFormula}), where the noise variable is already integrated out while the time step $\Delta \tau$ is still finite. In result, the SEO looses its mathematical meaning -- it is no longer a pullback averaged over the noise, but, rather, a result of formal manipulations with formulas.
Moreover, the error in the convergence of Eq.(\ref{SDE_discrete_time}) to Eq.(\ref{SDE}) mentioned in the previous paragraph, begins to interact with the noise yielding the $\alpha$-dependence of the SEO.

The so emerging ambiguity in the evolution operator is a general property of pathintegrals, not limited to stochastic dynamics, and also appears in quantum theory. It can only be removed by imposing additional conditions or principles. In quantum theory, the condition is the requirement for a Hermitian Hamiltonian, which is satisfied by the Weyl symmetrization rule corresponding to $\alpha=1/2$. In STS, the condition can be that the SEO matches the GTO (\ref{GTO}), which is also achieved at $\alpha=1/2$. In other words, only the Stratonovich interpretation provides SEO that matches GTO and, consequently, has a clear-cut mathematical meaning -- the pullback averaged over noise. 

Other interpretations differ only by the shifted flow vector field in Eq.(\ref{SEO_shifted}), which, however, does not carry any new mathematics. 
\footnote{{
It is worth noting here that it is commonly asserted in the literature that the Ito interpretation is distinguished from a mathematical standpoint because of its connection to the concept of martingale.}
\cite{Oks10} 
{This view originates from the observation that, at $\alpha=0$, the right hand side of Eq.}(\ref{SDE_discrete_time}) 
{depends only on $x_{k-1}$ (and $\xi_k$) but not on $x_{k}$. This is typically interpreted to mean that the Ito scheme “does not look into the future,” since $x_{k}$ depends solely on the previous value $x_{k-1}$ (and on $\xi_k$). As pointed out in Appendix A.1 of Ref.}
\cite{OvcEntropy}, 
{
however, the same property holds for all the other interpretations of stochastic dynamics as well. The variable $x_{k}$ is a unique function of $x_{k-1}$ and $\xi_k$ for any $\alpha$. For $\alpha \ne 0$, however, the unique function is given by Eq.}
(\ref{SDE_discrete_time})
{ only implicitly and one must solve for $x_k$ to express it explicitly in terms of $x_{k-1}$ and $\xi_k$. In other words, the fact that the right hand side of Eq}
.(\ref{SDE_discrete_time}) {at $\alpha=0$ does not depend on $x_{k}$ may facilitate the numerical implementation of temporal propagation, but carries no deeper mathematical significance. Particularly, stochastic evolution does not "look into the future" at any $\alpha$ and the following is always true for top differential forms (probability distributions):  $P(x,t) = \int d^{D}x' {\mathcal M}^{(D)}_{tt'}(x|x') P(x',t'), t'<t$, where ${\mathcal M}^{(D)}$ is the corresponding part of the GTO.
}
} Having multiple interpretations of SDEs is redundant. That said, alternative interpretations are relevant in the context of numerical implementations of SDEs, where different schemes may be preferred depending on computational context.

\section{Topological field theory and stochastic dynamics}
\label{Sec:STS_TFT}
The Parisi-Sourlas method is peculiar in that sense that its entire action is $Q$-exact as if it is a gauge-fixing of an empty theory. This is a definitive feature of cohomological TFTs. \cite{TFT_BOOK} As a TFT, STS has objects of topological nature. 
\subsection{Witten and Lefschetz indices}
The Witten index is one of such objects. Its topological character can be qualitatively understood as follows. As mentioned at the beginning of Sec.\ref{Sec:ParisiSourlas}, the Witten index is obtained by rewriting the noise partition function in terms of the dynamical variables of the DS. When done correctly, this procedure should yield an object which represents the noise partition function. However, since the noise carries no information about the DS dynamics, this object must be insensitive to perturbations or continuous deformations of the model. In other words, the object must be a topological invariant.

On a more rigorous level, the topological nature of the Witten index can be seen by noting, once again, that the gauge-fixing character of the action ensures that only solutions of the SDE contribute into pathintegral representation of $W$ in, say, Eq.(\ref{Witten_1}). Each solution provides either positive or negative unity, 
\begin{eqnarray}
&W = \Big\langle \iint_{p.b.c} J(\xi)\bigg(\prod\nolimits_\tau \delta ^D \bigg( \Big(\dot x(\tau) - {\mathscr F}(x(\tau),\xi(\tau))\Big) \dd\tau \bigg)\bigg)  {\mathcal D}x \Big\rangle_\text{noise} \\ 
&\nonumber = \left\langle \sum_{\dot x = {\mathscr F}(x,\xi)} J(\xi)/|J(\xi)| \right\rangle_\text{noise} = \langle I_N(\xi)\rangle_\text{noise},
\end{eqnarray}
where the absolute value of the Jacobian in the denominator is the result of the functional integration of the bosonic delta-functionals, and $I_N(\xi) = \sum_{\dot x = {\mathscr F}(x,\xi)} \operatorname{sign}J(\xi)$, is the index of the so-called Nicolai map. \cite{Nicolai1,Nicolai2} In our case, this is the map from the space of closed paths to the space of configurations of the noise making these paths solutions of the SDE, $\xi^a(x) = G^a_i(\dot x^i - F^i)/(2\Theta)^{1/2}$. The index of the map can be viewed as a realization of the Poincaré–Hopf theorem on the infinite-dimensional space of close paths with the SDE playing the role of the vector field and with the solutions of the SDE playing the role of the critical points with index $\text{sign}J(\xi) = \operatorname{sign } \det \delta \xi/\delta x$. $I_N(\xi)$ is a topological object independent of $\xi$. It equals its own stochastic average which, in turn, equals the Witten index.

There are other ways to see the topological character of $W$, with the most general mathematical framework related to this issue being the Mathai-Quillen formalism. \cite{BLAU199395} From the point of view of DS theory, the most interesting way to see its topological character is to integrate out all the fields but those on the first and last time slices (see Fig.\ref{Figure_3}). This leads to,
\begin{eqnarray}
    &W = \Big\langle \int \delta^D(M(\xi)_{t't}(x) -x)\delta^{D}(TM(\xi)_{t't}\chi - \chi) \dd^Dx \dd^D\chi\Big\rangle_\text{noise} \\ &\nonumber = \langle I_{L}(\xi)\rangle_\text{noise},    
    \;\text{ where } I_{L}(\xi) = \sum_{x=M(\xi)_{t't}(x)} \text{sign } \text{det} \big( TM(\xi)_{t't}(x) - \hat 1_{TX_x}\big) ,
\end{eqnarray}
is the Lefschetz index of $M(\xi)_{tt'}$, which is independent of the noise configuration and equals the Euler characteristic of $X$.
\subsection{Instantons and Morse-Smale dynamical systems}
\label{sec:InstantonsNMorseSmale}
Another class of objects of topological character are instantons or, more accurately, certain matrix elements on instantons. These objects are the reason why cohomological TFTs are identified sometimes as intersection theories on instantons. Instanton matrix elements is our next subject of interest.

Let us begin, however, by pointing out that from the physical point of view, instantons are the fundamental building blocks of transient dynamics in strongly nonlinear DSs. Earthquakes, solar flares, neuronal avalanches, and balloon popping are examples of instantons. Any given instance of transient dynamics, however, can be viewed as a composite instanton, i.e., a sequence of \emph{elementary} instantons. Examples of composite instantons include protein folding, the collapse of a building, or even the life circle of an organism, which can also be looked upon as a very complex composite instanton. Composite instantons may appear in response to quenches, i.e., sudden changes of conditions, where a DS is abruptly placed in an unstable position of its phase space and begins its evolution toward a stable attractor, as seen, for instance, in impact defragmentation. Another type of composite instantons is nonlinear dynamics induced by a slow change of external parameters, as in the crumpling paper or the Barkhausen effect.

From the mathematical point of view, instantons are transition processes between critical points or other invariant sets. As will be clear below, in this section we are talking about Morse-Smale DSs, the ones whose invariant sets are hyperbolic and have topologically well defined local stable or unstable manifolds. \cite{Morse_Smale_DS,book_hyperbolic_flows} Moreover, unlike antiinstantons (see Sec.\ref{Antiinstantons} below), instantons are not directly related to noise. Therefore, they can be considered in the deterministic limit, which we adopt here. 

A pathintegral representation of a matrix element on an instanton from a critical point $a$ to $b$ can be expressed as:
\begin{eqnarray}
&\langle b| \hat O | a\rangle = \iint_{x(\pm\infty)=x_{b,a}} O(x(0)\chi(0)) e^{( Q,\int_{-\infty}^\infty i\bar\chi (\dot x - F)\dd\tau )} {\mathcal D}\Phi,\label{Instanton_Introduction_0}
\end{eqnarray}
where $x_{a,b}$ are positions of the critical points such that $F(x_{a,b}) = 0$, and $O$ is an operator assumed, for simplicity, to depend only on $x\chi$ at $t=0$. In Eq.(\ref{Instanton_Introduction_0}), $\langle b|$ and $|a\rangle$ are the bra and the ket of the so-called local supersymmetric states of the corresponding critical points. These objects will be defined in the next section. In a meantime, let us use $\langle b|$ and $|a\rangle$ to avoid the necessity to introduce redundant notations.

The gauge-fixing character of the action of STS restricts the pathintegration only to the deterministic solutions that start at $a$ and end at $b$. There are infinitely many such solutions and their union forms an instanton manifold, $I_{ba}$, and $\text{dim }I_{ba} = \text{ind }a - \text{ind }b$, where the index of a critical point is the number of the unstable directions of the flow at this critical point. The points from the instanton manifold parametrise these solutions, $\dot x_{cl}(t,\sigma) = F(x_{cl}(t,\sigma))$, $x_{cl}(\pm\infty) = b,a;$ $\sigma \in I_{ab}$, where $\sigma$ are coordinates on $I_{ab}$ called instanton modulii. The differentiation of solutions with respect to the moduli yields zero modes of operator,
\begin{eqnarray}
    &\hat D(\sigma) = \dd/\dd t - TF(x_{cl}(t,\sigma));\; \hat D(\sigma) \partial x_{cl}(t,\sigma)/\partial \sigma = 0.  
\end{eqnarray}
These are the only zero-modes of $\hat D$ and $\hat D^\dagger$ (see, e.g., Ref.\cite{TFT_BOOK}). 

One can now introduce fluctuations around deterministic solutions as 
\begin{eqnarray}
    &x(t) = x_{cl}(t,\sigma) + \delta x'(t),
\end{eqnarray}
where the prime indicates that the fluctuations are restricted to directions transverse to the modulii. The corresponding decomposition for the superpartner $ \chi(t) = (\partial x_{cl}(t,\sigma)/\partial \sigma) \chi_\sigma + \chi'(t)$, where $\chi_\sigma = (Q, \sigma)$, $\chi'(t) = (Q,\delta x'(t))$. 

The instanton matrix element (\ref{Instanton_Introduction_0}) can now be expressed as 
\begin{eqnarray}
&\langle b| \hat O | a\rangle = \iint ( O(\sigma\chi_\sigma) + ... )  e^{( Q,\int_{-\infty}^\infty i\bar \chi \hat D (\sigma ) \delta x' \dd\tau)} {\mathcal D}\Phi' \dd^{D_I}\sigma \dd^{D_I}\chi_\sigma,\label{Instanton_2}
\end{eqnarray}
where $D_I=\text{dim} I_{ba}$, $O(\sigma \chi_\sigma) = O(x_{cl}(0,\sigma),(\partial x_{cl}(0,\sigma)/\partial \sigma) \chi_\sigma)$, $\Phi'$ denotes integration of all the fields except the instanton modulii and their superpartners, and the dots denote other terms that do not contribute, as will be pointed out shortly.

Integrating out the fluctuations in Eq.(\ref{Instanton_2}) yields a factor that equals unity (up to a sign), due to the cancellation of the bosonic and fermionic contributions, which is a very general principle in supersymemtric field theories, \cite{MirrorSymmetry} 
\begin{eqnarray}
    &\iint e^{( Q,\int_{-\infty}^\infty i\bar \chi \hat D (\sigma ) \delta x' \dd\tau )} {\mathcal D}\Phi' = \iint e^{\int_{-\infty}^\infty i B \hat D (\sigma ) \delta x' \dd\tau } {\mathcal D}B{\mathcal D}\delta x' \times \\
    & \times  \iint  e^{-\int_{-\infty}^\infty i\bar \chi \hat D (\sigma ) \chi' \dd\tau } {\mathcal D}\bar \chi{\mathcal D}\chi' =  \frac{ \text{det} \hat D'(\sigma)}{\big(\text{det}\big(\hat D'(\sigma)\hat {D'}^\dagger(\sigma)\big)\big)^{1/2}} = \text{sign det} \hat D'(\sigma) ,\nonumber
\end{eqnarray}
where $\hat D'(\sigma)$ is a restriction of $\hat D(\sigma)$ on all but zero modes. 

One of the basic rules of integration over fermionic fields is the requirement that every fermionic differential must be matched by the corresponding fermion according to $\int \dd \chi =0, \int \chi \dd\chi = 1$. In application to Eq.(\ref{Instanton_2}), this means that $O$ must provide all the $\chi_\sigma$'s to match, $\dd^{D_I}\chi_\sigma$. This provision is accomplished by the first term in the resolution of $O$ in the parentheses in the r.h.s. of Eq.(\ref{Instanton_2}), while the other terms do not contribute to the matrix element, assuming that the degree of $O$ equals the dimensionality of the instanton manifold. Thus, we arrive at
\begin{eqnarray}
&\langle b| \hat O| a\rangle = \int O(\sigma \chi_\sigma)  \dd\sigma \dd\chi_\sigma = \int_{I_{ab}} O,\label{Instanton_Introduction}
\end{eqnarray}
where $O$ in the r.h.s. is understood as a differential form on $I_{ba}$, which can be interpreted as a pullback of $O$ from the space of paths to $I_{ba}$ provided by $x_{cl}$.
\begin{figure}[t]
\centerline{\includegraphics[width=0.7\linewidth]{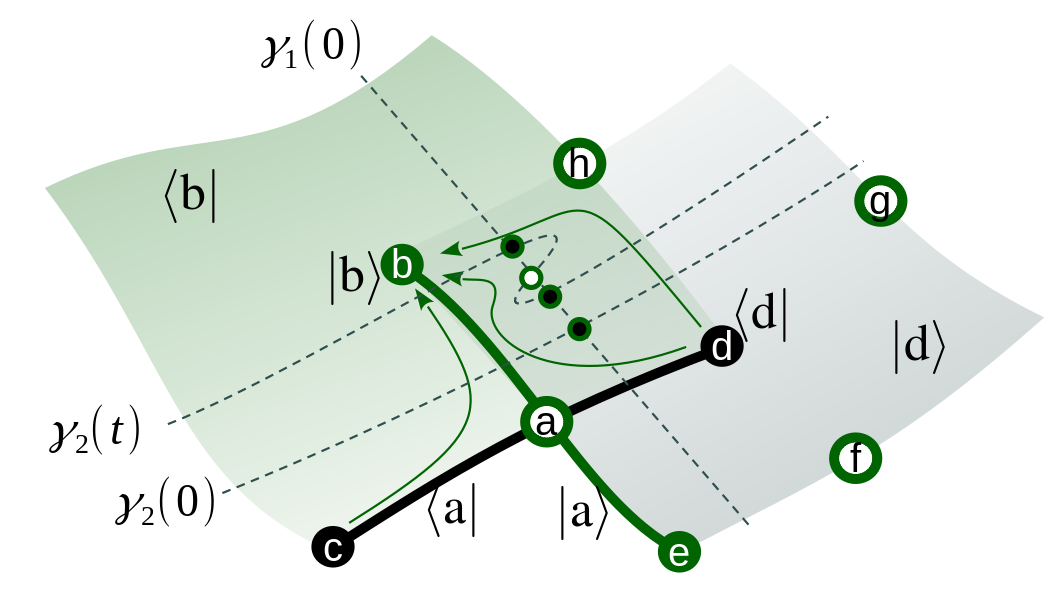}}
\caption{\label{Figure_Smale_Complex} An example of a Morse-Smale flow (thin green arrowed  curves). The filled green circles (b, e) represent minima (index 0), hollow circles (a,f,g,h) correspond to saddles (index-1), and filled-black circles (c,b) denote index-2 critical points. The bras/kets of the local supersymmetric states of the Morse-Smale-Bott-Witten complex are Poincare duals of the local stable/unstable manifolds. For example, $\langle a| = p(S_a)$ and $|a\rangle=p(U_a)$ are narrow distributions on $S_a=(cad)$ and $U_a=(eab)$, respectively, with fermions in transverse directions, whereas $\langle b|$ and $|d\rangle$ are constant functions over the green and gray regions, respectively. The dashed curves represent the one-parameter families of 1-dimensional manifolds, $\gamma_1(t), \gamma_2(t)$, obtained by the flow-defined diffeomorphisms, $\gamma_1(t) = M_{t0}(\gamma_1)$. Their Poincare duals can be used to construct, e.g, the matrix element, $\langle b| \hat p(\gamma_1(t))\hat p(\gamma_2(0))|d\rangle =1$, which represents the intersection number of $\gamma$'s on the instanton manifold, $I_{bd}  = S_b \cap U_d =(bhda)$. The matrix element is independent of $t$'s because the intersection points (dis)appear in pairs with opposite orientations (white and black filled circles).}
\end{figure}
\subsubsection{Morse-Smale-Bott-Witten complex}
To examine the instanton matrix element in the operator representation, let us define the local supersymmetric states (LSS), whose notation we already introduced in the previous section:
\begin{eqnarray}
&\langle x\chi|a\rangle   = \iint_{{x(-\infty)=x_a} \atop {x\chi(0)=x\chi}} e^{( Q,\int_{-\infty}^0 i\bar\chi(\dot x - F)\dd\tau)} {\mathcal D}\Phi,\\
&\langle b | x\chi\rangle = \iint_{{x(+\infty)=x_b} \atop {x\chi(0)=x\chi}} e^{( Q,\int_0^{+\infty} i\bar\chi(\dot x - F)\dd\tau)} {\mathcal D}\Phi.
\end{eqnarray}
Given the gauge-fixing nature of the action limiting the pathintegration to deterministic solution of the flow, it follows that $|a\rangle$ is non-zero only for points that flow to $a$ in the $t=-\infty$ limit, that is, for points on the local unstable manifold of $a$: $U_a$.\footnote{The local stable and unstable manifold of a critical point are defined as the set of points that asymptotically flow toward the critical point in the infinite future and past, respectively.} Similarly, $\langle b|$ is non-zero only on the local stable manifold of $b$: $S_b$. That the intersection of the local stable and unstable manifolds is the instanton manifold, $I_{ba}= U_a \cap S_b$, is a well known set-theoretic result of DS theory. From the algebraic point of view, however, the fermionic structure of the LSSs is also important. 

To determine the fermionic content of a LSS, let us consider the simple case of a Langevin SDE on $X=\mathbb{R}$ with $F=-U', U = \omega x^2/2$ and an additive noise. Its SEO can be expressed as,
\begin{eqnarray}
    \hat H = e^{-U/2\Theta} \hat H_W e^{U/2\Theta}, \;   \hat H_W =\hat H_W^\dagger = \Theta [\hat d_W, \hat d_W^\dagger],
\end{eqnarray}
where $\hat d_W = \chi( - U'/2\Theta + \partial/\partial x )$, $\hat d_W^\dagger = (\partial/\partial\chi)( - U'/2\Theta - \partial/\partial x )$, and  $\hat H_W$ can be identified as the Hermitian Hamiltonian of a 1D supersymmetric harmonic oscillator. \cite{Junker1996} $\hat H$ and $\hat H_W$ are related via a similarity transformation so that their eigensystems are identical up to this transformation. In terms of $\hat H_W$, the ket and the bra of the zero-eigenvalue supersymmetric ground state are, respectively, $\psi_W = \chi e^{-|\omega|x^2/2\Theta}$ and $\bar\psi_W = e^{-|\omega|x^2/2\theta}$ for $\omega>0$, and $\bar\psi_W \leftrightarrow \psi_W$ for $\omega<0$. In terms of $\hat H$, for $\omega>0$, the ket of the ground state is $\chi e^{-|\omega|x^2/\Theta}$, while the bra is a constant function. For $\omega<0$, the roles of bra and ket are reversed.

This example can be easily generalized to the multiple-variable non-degenerate critical point of a gradient flow. In appropriate coordinates, $U = \sum_i \omega_i(x^i)^2/2, \omega_i\ne0, i = 1...D$. Both the bra and the ket of the LSS factorize, with each coordinate contributing a factor from the 1D Langevin case above. In the deterministic limit ($\Theta\to0$), the ket of the LSS is a narrow distribution with fermions in the stable directions ($\omega_i>0$) and a constant function without fermions in the unstable directions ($\omega_i<0$). The situation is reversed for the bra, which is a narrow distribution in the unstable directions and a constant function in the stable directions.

A natural generalization of the previous example is that the ket and the bra of the LSS associated with an isolated critical point, $a$, are the Poincare duals, $p(U_a)$ and $p(S_a)$, of the local unstable and stable manifolds, respectively. \footnote{By definition, the Poincare dual, $p(Z)$, of a submanifold $Z$ is a $\delta$-distribution on $Z$ with differentials in transverse directions. Its key property is $\int_{Z} \psi = \int_X p(Z)\wedge \psi, \forall \psi$. For example, for a co-dimension $1$ hyperplane, $\gamma_{0} = \{ x\in \mathrm{R}^D | x^i = x_0 \}$, the Poincare dual $p(\gamma_0) = \delta(x^i-x_0)\dd x^i$. } To see that this is indeed so, recall that in the deterministic limit, the SEO consists only of the Lie derivative along $F$ and the Poincare duals of (un)stable manifolds of the flow lie in its kernel, that is, they are zero-eigenvalue LSSs of the evolution operator. 

In terms of Poincare duals, the matrix element (\ref{Instanton_Introduction}) can be expressed as
\begin{eqnarray}
    \langle b| \hat O | a\rangle = \int_X p(S_b) \wedge O \wedge p(U_a).
\end{eqnarray}
Being supersymmetric states, the LSSs are $\hat d$-closed, $\hat d p(U_a)=0$. This property, however, holds only locally in the vicinity of the critical point, justifying the term "local supersymmetric states". From the global perspective, LSSs provide an algebraic representation of the Morse-Smale-Witten complex, with $\hat d$ serving as the algebraic counterpart of the boundary operator. For instance, the action of $\hat d$ on some of the LSSs in Fig.\ref{Figure_Smale_Complex} is,  
\begin{eqnarray}
\hat d |a\rangle = |b\rangle - |e\rangle,\; \hat d |d\rangle = |a\rangle + |h\rangle - |g\rangle - |f\rangle, ...
\end{eqnarray}
This framework should extend naturally to nontrivial invariant sets via the Morse-Bott approach (see, \emph{e.g.}, Ch.10 of Ref.\cite{MirrorSymmetry}). Each de Rham cohomology class of an invariant set must provide one (global in terms of the invariant set) supersymmetric state (see Sec.\ref{Sec:susySinglets}). These may serve as additional factors for the Morse-Smale-Witten LSSs leading to a generalized structure that could be termed the Morse-Smale-Bott-Witten complex. While the present author is unaware of a rigorous establishment of this extension, he finds it natural to expect its validity for Morse-Smale DSs, whose Morse-Smale complex is topologically well-defined and structurally stable. \cite{Morse_Smale_DS,book_hyperbolic_flows}

\subsubsection{Intersections on instantons}
As mentioned earlier, certain matrix elements on instantons in TFTs are topological invariants. \cite{Frenkel2007215} These are matrix elements of $Q$-closed operators on LSSs. In the class of models considered here, the simplest topological invariants of this type can be defined as follows.

Consider a set of submanifolds in $X$, $\{\gamma_\alpha | \alpha = 0,1,... \}$. Their Poincare duals, $p(\gamma_\alpha)$, satisfy the relation,
\begin{eqnarray}
(Q,  p(\gamma_i)) = [\hat d, p(\gamma_i) ] = p(\partial \gamma_i),
\end{eqnarray}
where $\partial\gamma_i$ is the boundary of $\gamma_i$. The duals $p(\gamma)$'s are $Q$-closed if $\gamma$'s are either boundaryless or if their boundaries lie outside the domain of $X$ under consideration. 

Consider also the following matrix elements,
\begin{eqnarray}
&\iint_{x_{\pm\infty}=x_{b,a}} p(\gamma_k)(t_k) ... p(\gamma_1)(0) e^{( Q,\int_{-\infty}^\infty i\bar\chi (\dot x - F)\dd\tau )} {\mathcal D}\Phi \\ \nonumber &= \langle b |\hat p(\gamma_k)(t_k)...\hat p(\gamma_1)(t_1)|a \rangle,
\end{eqnarray}
where without loss of generality we assume $t_k>...>t_1$, and
\begin{eqnarray}
&\hat p(\gamma_i)(t_k) = e^{t_i\hat H}\hat p(\gamma_i) e^{-t_i\hat H} = \hat M^*_{t_i0}(p(\gamma_i)) = p( \gamma_i(t_i)),\label{no_label_eq}
\end{eqnarray}
are the corresponding operators in the so-called Heisenberg representation, $\hat M^*_{t_i0}$ is the pullback induced by $M_{t_i0}$, and $\gamma_i(t_i) = M_{t_i0}(\gamma_i)$ is the manifold obtained from $\gamma_i$ through the diffeomorphism, $M_{t_i0}$. The validity of Eq.(\ref{no_label_eq}) follows from the observation that in the deterministic limit the SEO is just the flow along $F$. 

It is now clear that the above matrix element represents the intersection number,
\begin{eqnarray}
&\int_{I_{ba}} p(\gamma_k(t_k)) \wedge ... \wedge p(\gamma_1(t_1)) = \sum_{x\in I_{ba}\cap  \gamma_k(t_k) \cap ...\cap  \gamma_1(t_1) } (\pm),\label{Intersection}
\end{eqnarray}
where the signs account for the mutual orientation of $\gamma$'s at the intersection points, and it is assumed that the sum of codimensions of $\gamma$'s equals the dimensionality of $I_{ba}$. The topological invariance of Eq.(\ref{Intersection}) is its independence of $t's$: the flow-induced diffeomorphism acting on any $\gamma$ does not alter the intersection number because the intersection points (dis)appear in pairs with opposite orientations (see Fig.\ref{Figure_3}).

\subsubsection{Antiinstantons}
\label{Antiinstantons}
The qualitative discussion in Sec.\ref{Sec:Phase_Diagram} below relies partly on the concept of antiinstantons. These are the time-reversed instantons of motion against the flow. Unlike instantons, antiinstantons are only possible with the assistance from the noise. As a result, their matrix elements contain exponentially weak Gibbs factors vanishing in the deterministic limit. This can be seen, for example, in the Langevin SDEs where antiinstantonic matrix elements are related to their instantonic counterparts as:
\begin{eqnarray}
  \langle a| \hat {O}^\dagger | b \rangle = e ^{-2(U(x_b)-U(x_a))/\Theta} \langle b| \hat O | a \rangle,
\end{eqnarray}
where $U$ is the Langevin potential defining the flow $F=-\partial U$, and $\hat{O}^\dagger$ is the conjugate of $\hat O$, obtained by the substitutions $\chi \leftrightarrow \bar\chi $ and $B\to B + 2F/\Theta$. \cite{TFT_BOOK} 

\begin{figure}[t]
\centerline{\includegraphics[width=0.65\linewidth]{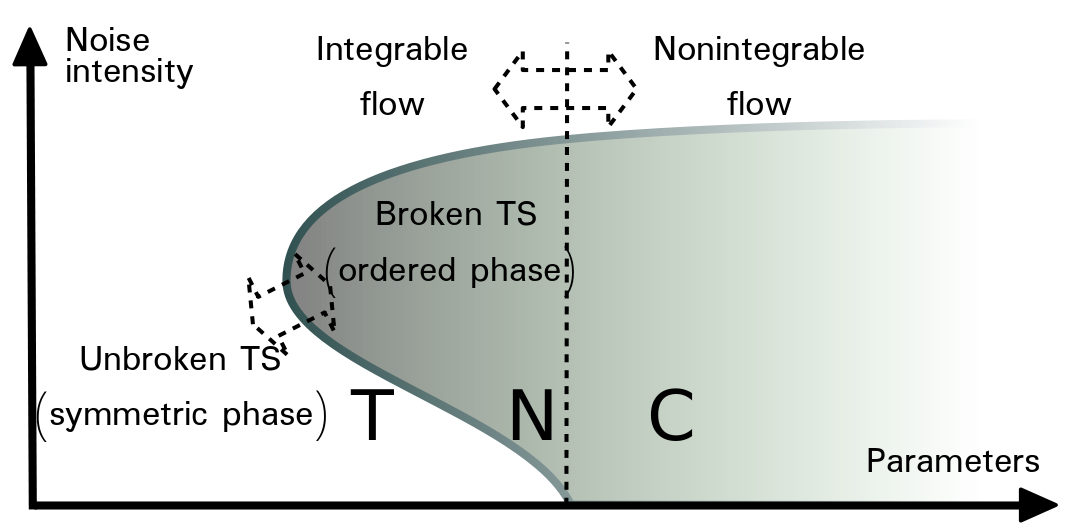}}
\caption{\label{Phase_Diagram} Stochastic DSs can be classified based on two key factors: \emph{(i)} whether the topological supersymmetry (TS) is spontaneously broken (ordered phase) or unbroken (symmetric phase) and \emph{(ii)} whether the flow vector field is integrable or non-integrable and/or chaotic. The symmetric phase with unbroken TS is labeled as T. The ordered phase with non-integrable flow (C-phase) is a stochastic generalization of the deterministic chaos, where the TS breaking is caused by the nonintegrability of the flow. The ordered phase with integrable flow (N-phase) can be identified as the noise-induced chaos, where the dynamics is dominated by noise-induced instantons. The instantons vanish in the deterministic limit, causing the N-phase to collapse onto the boundary of the C-phase. As noise intensity increases, the TS must eventually be restored disregard of the properties of the flow, as the GTO/SEO becomes dominated by the Laplacian, which alone does not break TS.}
\end{figure}

\section{Self-sustained dynamics}
\label{Sec:Phase_Diagram}
As discussed at the beginning of Sec.\ref{Sec:STS_TFT}, transient dynamics is the concept representing a strongly nonlinear dissipative DS on its way to a stable attractor. In contrast, sustained dynamics refers to the state of the DS after infinitely long evolution unperturbed by external influence other than the stochastic noise. While transient dynamics is associated with instantons and local supersymmetric states, sustained dynamics is described by the global ground state. The very concept of the spontaneous symmetry breaking pertains to the global ground state.

\subsection{Global ground state}
The global ground state is a part of the definition of the generating functional, $G(\eta)$, -- a functional of external probing fields, $\eta$, introduced into the SDE, $F\to F(\eta)$, to explore the system's response to external influence. It is understood that $G(\eta)$ must be constructed from the SEO, which is the most important object in the theory and which is also a functional of $\eta$ in the presence of the probing fields, $\hat M _{T/2, -T/2}(\eta)$. 

The sharp trace of the SEO is unsuitable as a generating functional because, as a topological invariant, it is insensitive to the external perturbations. The ordinary trace of the GTO is a better alternative. However, it is still not good enough: in DSs with the broken pseudo-time-reversal symmetry (type-c spectra in Fig.\ref{Figure_2}), such a generating functional would exhibit undesirable oscillatory behavior in the long-time limit. The optimal choice for the generating functional is,
\begin{eqnarray}
&G(\eta) = -\log \lim_{T\to\infty} e^{H_g T} \langle g | \hat M _{T/2, -T/2}(\eta) | g \rangle,
\end{eqnarray}
where the factor $e^{H_g T} = 1/\langle g | \hat M _{T/2, -T/2}(0) | g \rangle$ is introduced to remove the unimportant infinite constant and $| g \rangle$ is the global ground state, i.e., one of the eigenstates with the smallest real part of the eigenvalue, $\text{Re } H_g=\min\nolimits_{\alpha}\text{Re }H_\alpha$. This criterion for the eigenvalue of the ground state ensures the stability of the response. 

The functional dependence of $G(\eta)$ on the probing fields characterizes the ground state’s response to external perturbations. Since this response involves transitions to all other eigenstates, the choice of a particular eigenstate as the ground state among a few fastest growing eigenstates does not impose any limitations onto the so-obtained description of the system's dynamics.

\subsubsection{Spontaneous topological supersymmetry breaking}
When the TS is broken spontaneously, an eigenstate with the same eigenvalue as the ground state exists, making the ground state easily "excitable". This effortless excitability is a predecessor of the Goldstone theorem, which states that in higher-dimensional models, where basespaces have dimensions other than time, a gapless excitation must exist. These excitations, known as goldstinos, mediate long-range responses and may provide a qualitative explanation for the ubiquity of power-law correlations in chaotic dynamics, commonly referred to as 1/f noise.

A rigorous theoretical explanation of 1/f noise remains an open question. However, there is a quantitative argument supporting this claim. This argument applies to higher-dimensional models with a long-range dynamics of Lorenzian type such as the one discussed in Ref.\cite{OVCHINNIKOV2024114611}. In such models, the large-scale dynamics must be scale invariant and the corresponding effective field theory (EFT) \cite{RFT_SSB_book} must be a conformal field theory (CFT). \cite{ginsparg1988appliedconformalfieldtheory} \footnote{There are also reasons to believe that the EFT is not only a CFT but also a TFT.\cite{ovchinnikov2025topologicalnaturebutterflyeffect}} As a CFT, the EFT must possess a set of primary local fields, $\hat O_i(r), i=1,..,N$, where $r$ is a basespace point, such that, $\langle g | \hat O_i(r) | g \rangle=0$, and
\begin{eqnarray}
    &\langle g | \hat O_i(0)\hat O_i(r)| g \rangle =1/|r|^{2\Delta_i}. \label{Primiries}
\end{eqnarray} 
where $\Delta$'s are the conformal weights of the primary fields. Furthermore, by the so-called operator-state correspondence in CFTs, any local operator, $\hat O(r)$, can be resolved as
\begin{eqnarray}
    &\hat O(r) = \sum\nolimits_{i} c_{i}\hat O_i(r) + ... ,\label{OpStateCorr}
\end{eqnarray}
where the omitted terms represent descendant fields, which are subdominant in the long-wavelength limit as they have higher conformal weights. Eqs.(\ref{Primiries}) and (\ref{OpStateCorr}) lead to the conclusion that in the long-wavelength limit, correlators of a wide class of observables must be a power law   
\begin{eqnarray}
   &\langle g | \hat O(0) \hat O(r)|g\rangle\big|_{|r|\to\infty} = c_{i_s}^2/|r|^{2\Delta_{i_m}} + ..., \label{Mellin}
\end{eqnarray}
where $i_s$ is the index of the primary field with the smallest conformal weight.

\subsubsection{Spontaneously pseudo-time-reversal symmetry breaking} 
\label{EtaTSymmetry}
Another interesting situation arises when $H_g$ is complex. In this case, the "relative" eigenvalues, $\Delta H_\alpha = H_\alpha-H_g$, of the low-lying eigenstates -- which govern the long-range behavior -- are no longer real or complex conjugate pairs, signaling the breakdown of pseudo-time-reversal symmetry. In the context of kinematic dynamo theory, a complex $H_g$ corresponds to the rotation of a growing galactic magnetic field. \cite{Torsten} The broader implications of a complex $H_g$ is an open question.

\subsection{Phase Diagram} 
Our final topic of interest is the internal structure of the global ground state. A thorough analysis of this problem may take time to fully resolve. At present, only a qualitative understanding is available, which is expected to evolve with future research, leading to refinements and deeper insights. Despite its preliminary nature, this understanding provides valuable insights, making it a topic worthy of a brief discussion.

\subsubsection{Integrable flows and unbroken TS}
In deterministic Morse-Smale DSs (see Sec.(\ref{sec:InstantonsNMorseSmale})), the local (un)stable manifolds with boundaries can be glued together into topologically well-defined global (un)stable manifolds that form foliations of $X$. The existence of the topologically well-defined global unstable manifolds is actually a criterion for a flow to be identified as integrable in the sense of DS, \emph{i.e.}, the ones satisfying Frobenius' theorem. \cite{Intro_Smooth_Manifolds} In such situations, and according to the discussion in the previous section, the Poincare duals of the global (un)stable manifolds are the global ground states. These ground states are supersymmetric because they are $\hat d$-closed, which is the algebraic version of the statement that the global (un)stable manifolds have no boundaries. 

\subsubsection{Non-integrable flows and TS breaking}
For chaotic or non-integrable deterministic DSs, the topologically well-defined global (un)stable manifolds do not exist. For example, in topological theory of chaos, \cite{Gilmore_book} the global unstable manifolds are approximated by branched unstable manifolds. The branched manifolds have self-intersections so that they are not topological manifolds. As a result, the Poincare dual of such a branched (un)stable manifold, which is supposed to be the ground state of the model, cannot be $\hat d$-closed. In other words, the ground state is not supersymmetric for chaotic deterministic flows and TS is broken spontaneously.

\subsubsection{Noise-induced chaos and instanton condensation}
For an integrable flow, whose Morse-Smale complex is stable with respect to weak perturbations, introduction of a sufficiently weak noise must not break TS. the noise will only broaden the ground state in transverse directions of the global (un)stable manifold, making it -- in the first approximation -- a narrow supersymmetric Gaussian distribution. In high-energy theory terminology, this implies that the TS is unbroken at the Gaussian level. Higher-order perturbative corrections must not qualitatively change this picture because supersymmetries are robust symmetries: if they are not broken on the Guassian level, then higher-order perturbative corrections cannot break them either -- the well-known absence of the supersymmetry breaking anomaly. \cite{Bertlmann1996,AlvarezGaume1984} 

However, sufficiently strong noise can break the TS of the integrable flows through a mechanism distinct from perturbative corrections -- the condensation of the noise-induced antiinstantons and instantons matching them, or, simpler, the condensation of the noise-induced instantons. \footnote{Instantons is a reliable mechanism of the spontaneous supersymmetry breaking in high-energy physics. \cite{DynSusyBrWitten}} When this happens, the resulting dynamics should look as an endless sequence of the noise-induced instantons interacting with each other. Moreover, certain characteristics of instantons, such as their statistical properties -- exemplified by the power-law distribution of earthquakes, solar flares, neuroavalanches, etc. -- should reflect the long-range nature of chaos, in accordance with the Goldstone theorem. Importantly, the mere existence of instantons in a nonlinear model does not necessarily break TS. \footnote{For example, in Langevin SDEs with multiple minima of the Langevin potential, instantons exist, yet TS is never broken because the eigenvalues of the SEO are real and non-negative.} Noise-induced instantons can only facilitate TS breaking in flows that are already close to being chaotic (see Fig.\ref{Phase_Diagram}).

It now follows that this type of dynamics, which can be called the noise-induced chaos, must collapse onto the boundary of deterministic chaos in the deterministic limit. This conclusion follows from two observations: \emph{(i)} noise-induced chaos disappears in the deterministic limit, just like the anti-instantons that underlie it, and \emph{(ii)} the TS-breaking phase transition must form a continuous curve that terminates precisely at the edge of deterministic chaos as the deterministic limit is approached. In this way (see Fig.\ref{Phase_Diagram}), STS provides a theoretical framework for understanding the phenomenon known as the "border of chaos." \cite{DynamicalComplexity,Crutchfield}

\section{Conclusion}
\label{Sec:Conclusion}
In this paper, we discussed key aspects of the recently established connection between dynamical systems theory and cohomological topological field theories, a framework that can be referred to as the supersymmetric theory of stochastic dynamics. We demonstrated that this approach is an algebraic dual to the set-theoretic framework of dynamical systems theory. The added algebraic structure reveals the presence of the topological supersymmetry in all stochastic models and enables the stochastic generalization of concepts traditionally associated with deterministic dynamics. Namely, the Morse-Smale complex of local unstable manifolds, strange attractors, and chaoticity of a deterministic flow correspond, on the side of stochastic dynamics, to the Morse-Smale-Bott-Witten complex of local supersymmetric states, global non-supersymmetric ground states, and the spontaneous breakdown of topological supersymmetry, respectively.

{From a practical standpoint, STS is particularly interesting because it provides an explanation for the experimental signature of chaotic dynamics known as 1/f noise. Numerous attempts have been made to account for this phenomenon; the most prominent among them is perhaps the concept of self-organized criticality — the idea that some mysterious force fine-tunes stochastic dynamical systems to the transition into chaos.}\cite{Bak87} {To the best of the present author’s knowledge, however, STS is the only framework that offers a theoretically rigorous explanation of 1/f noise.}

Beyond that, STS has yielded fresh insights into the competing definitions of chaos and the various interpretations of stochastic dynamics, offering a theoretical understanding of behavior at the so-called “edge of chaos.” Its potential implications, however, extend even further. More broadly, mathematical physics today is divided into two major camps: quantum and classical. The gap between them is substantial, making collaboration between, for example, dynamical systems theorists and string theorists challenging. The supersymmetric theory of stochastic dynamics has the potential to bridge this divide by unifying concepts and providing a shared mathematical framework, fostering collaboration and accelerating progress in both areas.

\section*{Acknowledgments}
The author acknowledges the initial support from DARPA BAA "Physical Intelligence" and extends special gratitude to Kang L. Wang for his pivotal role in enabling this work. Sincere thanks are also extended to Massimiliano Di Ventra, Torsten A. Ensslin, Robert N. Schwartz, Savdeep S. Sethi, Gabriel A. Weiderpass, Ben Israelii, Daniel Toker, Skirmantas Janusonis, Dmitri A. Riabtsev, Cheng-Zong Bai, and Eugene Ingerman, all of whom positively influenced this work.

\section*{Declaration of conflict of interest}
The author declares no conflicts of interest.

\section*{Data availability statement}
No datasets were generated or analyzed in this work.

\bibliographystyle{amsplain}
\bibliography{my-bibliography}

\end{document}